\documentclass[a4paper,11pt]{article}
\usepackage{jheppub}
\usepackage{lineno}
\usepackage{xcolor}
\usepackage{tabularx}
\usepackage{bm}
\usepackage{graphicx}
\usepackage{subcaption}
\usepackage{booktabs} 
\usepackage{amsmath}   
\usepackage{gensymb}         
\usepackage{siunitx}        
\usepackage[normalem]{ulem}

\renewcommand{\figurename}{Fig.}
\title{\boldmath Vertex reconstruction in the TAO experiment}

\author[a, c, d]{Hangyu Shi, }
\author[a, c, d, *]{Jun Wang, } 
\author[e, *]{Guofu Cao, }
\author[a, b, c, d, *]{Wei Wang, }
\author[a, c, d]{and Yuehuan Wei}

\affiliation[a]{Sino-French Institute of Nuclear Engineering and Technology, Sun Yat-sen University, Zhuhai, 519082, China.}
\affiliation[b]{School of Physics, Sun Yat-sen University, Guangzhou, 510275, China.}
\affiliation[c]{GuangDong Engineering Technology Research Center of Nuclear Safety and Emergency Technology, Zhuhai, 519082, China.}
\affiliation[d]{Research Center for Nuclear Technology and Applications of SYSU, Zhuhai, 519082, China.}
\affiliation[e]{Institute of High Energy Physics, Chinese Academy of Sciences, Beijing, 100049, China.}
\footnotetext[1]{Corresponding author.}

\emailAdd{shihy35@mail2.sysu.edu.cn}
\emailAdd{wangj933@mail.sysu.edu.cn}
\emailAdd{caogf@ihep.ac.cn}
\emailAdd{wangw223@mail.sysu.edu.cn}
\emailAdd{weiyh29@mail.sysu.edu.cn}

\abstract{The Taishan Antineutrino Observatory (TAO) is a tonne-scale gadolinium-doped liquid scintillator satellite experiment of the Jiangmen Underground Neutrino Observatory (JUNO). 
It is designed to measure the reactor antineutrino energy spectrum with unprecedented energy resolution, better than 2\% at 1 MeV. 
To fully achieve its designed performance, precise vertex reconstruction is crucial. 
This work reports two distinct vertex reconstruction methods, the charge center algorithm (CCA) and the deep learning algorithm (DLA). 
We describe the efforts in optimizing and improving these two methods and compare their reconstruction performance. 
The results show that the CCA and DLA methods can achieve vertex position resolutions better than 20 mm (bias $<$ 5 mm) and 12 mm (bias $<$ 1.3 mm) at 1 MeV, respectively, fully meeting the requirements of the TAO experiment. 
The reconstruction algorithms developed in this study not only prepare the TAO experiment for its upcoming real data but also hold significant potential for application in other similar experiments.
}


\begin{document}
\maketitle
\flushbottom

\section{Introduction}
\label{sec:intro}

Reactor neutrino experiments such as RENO~\cite{RENO:2015ksa}, Daya Bay~\cite{DayaBay:2015lja}, Double Chooz~\cite{DoubleChooz:2015mfm}, NEOS~\cite{NEOS:2016wee}, and STEREO~\cite{STEREO:2022nzk} have revealed systematic deviations between the measured neutrino energy spectra 
and the predictions from the benchmark Huber-Mueller model~\cite{Huber:2011wv,Mueller:2011nm} currently adopted in neutrino analyses. 
These deviations include the reactor antineutrino anomaly (RAA)~\cite{Mention:2011rk}, characterized by an approximately 6\% flux deficit, and are accompanied by a distinctive spectral excess around 5 MeV (the so-called 5 MeV bump)~\cite{Seo:2014xei,Huber:2016xis}. 
If these predicted spectra were directly implemented as input for reactor neutrino experiments employing the inverse beta decay (IBD) reaction or coherent elastic neutrino-nucleus scattering (CE$\nu$NS), such as JUNO~\cite{JUNO:2024jaw} and RELICS~\cite{RELICS:2024opj}, it could introduce non-negligible spectral uncertainties that might compromise their sensitivity to their own primary physics goals. To mitigate potential model dependencies due to the fine structure of the reactor antineutrino spectrum, the JUNO collaboration has deployed the Taishan Antineutrino Observatory (TAO, also known as JUNO-TAO)~\cite{JUNO:2020ijm} as a dedicated satellite detector positioned at a baseline of 44 meters from one of the operational reactor cores at the Taishan Nuclear Power Plant. 

The central detector (CD) of TAO is designed to have a tonne-scale fiducial mass of Gadolinium-doped Liquid Scintillator (GdLS), featuring the same hydrogen atom mass fraction and abundance as JUNO~\cite{JUNO:2024jaw} and achieving over 93.5\% coverage of photon sensors and an energy resolution better than 2\% at 1 MeV~\cite{JUNO:2020ijm}. 
It will realize a neutrino event rate of approximately 1000 per day in the fiducial volume, which is more than 20 times higher than that of JUNO with selection efficiency included~\cite{JUNO:2024jaw}. 
Based on these specifications, the TAO experiment can precisely measure the reactor antineutrino energy spectrum with sub-percent energy resolution across the key energy ranges of interest. 
Consequently, not only does it provide a reference neutrino energy spectrum for current reactor neutrino experiments such as JUNO, but it also performs benchmark measurements to test nuclear databases, searches for light sterile neutrinos with a mass scale of around 1 eV, and validates detector technologies for reactor monitoring and safeguard applications, among other purposes, thereby establishing itself as a multirole experimental platform.

To achieve these multiple objectives, precise vertex reconstruction of detected events is essential. It allows for an adequate fiducial volume cut that suppresses backgrounds originating outside the target volume. 
Moreover, the vertex resolution directly influences the correction of spatial non-uniformities in the detector response, which in turn affects the accuracy of energy reconstruction and the control of systematic uncertainties in reactor neutrino spectrum measurements \cite{Xu:2022mdi}.
The TAO experiment has established a stringent target for its vertex reconstruction algorithm, requiring spatial resolution and bias to be better than 
5 cm~\cite{JUNO:2020ijm}.

The vertex reconstruction algorithms in liquid scintillator detectors now include both traditional and machine-learning approaches. 
In traditional vertex reconstruction, certain approaches utilize the charge of the pulses detected by the photomultiplier tube (PMT) or the time-of-flight of the photons to formulate analytical functions for directly computing the event vertex, such as the charge center algorithm (CCA) and the charge-pattern templates algorithm employed in the Daya Bay experiment~\cite{DayaBay:2012aa,DayaBay:2016ggj}, 
as well as the time-based algorithm used in the JUNO experiment~\cite{Li:2021oos}. Alternatively, other approaches construct likelihood functions based on the expected distributions of charge and/or time, deriving the event vertex by minimizing these functions. For example, the time likelihood algorithm was applied in the  Kamland~\cite{KamLAND:2002uet} and Borexino~\cite{Borexino:2008gab} experiments, while the charge-and-time combined likelihood algorithms have been developed in the Double Chooz~\cite{DoubleChooz:2012gmf}, JUNO~\cite{Liu:2018fpq, Huang:2022zum}, and TAO~\cite{Liu:2024cxo} experiments.

Machine learning approaches, especially deep learning algorithms (DLAs), have increasingly proven valuable in event vertex reconstruction in various high-energy and neutrino physics experiments because they can improve the accuracy and resolution of event reconstruction compared to traditional approaches. 
For instance, the JUNO experiment has applied convolutional neural network (CNN) and graph neural network (GNN) models to event vertex and energy reconstruction tasks~\cite{Li:2022gpb,Qian:2021vnh,Li:2022tvg}, with promising results based on simulation studies. 
Training on extensive simulated event data allows these deep learning models to learn complex nonlinear correlations between input variables, such as charge and timing data collected by PMTs, 
and the actual vertex positions, enabling accurate and rapid event vertex prediction even under challenging conditions. Furthermore, they facilitate real-time event processing, a vital aspect for experiments requiring high throughput.

In the case of small-scale detectors like TAO, the time resolution of 8 ns, due to the design of the electronics system~\cite{Venettacci:2024hwy}, is insufficient to resolve photon time-of-flight differences. This limitation hinders the construction of precise time templates. Instead, CCA relies solely on the spatial distribution of charge to efficiently compute the charge-weighted centroid with low algorithmic complexity and high computational efficiency, making it ideal for reliable and rapid reconstruction of large event sets. DLA eliminates the need for constructing templates by directly learning nonlinear mappings from the charge and time data, achieving exceptional reconstruction accuracy. Its inherent parallel processing capabilities enable large-scale data handling, making it a powerful complement to traditional approaches. Therefore, this paper focuses on CCA and DLA and systematically investigates the optimization strategies and performance characteristics of the two algorithms, proposing the corresponding high-precision vertex reconstruction solutions for the TAO experiment and ensuring robust preparation for the upcoming data acquisition phase.

The remainder of this paper is organized as follows. Section \ref{sec:TAO detector} provides an overview of the core detection framework of the TAO detector. Section \ref{sec: CCA} details the principles and enhancements of the CCA. In Section \ref{sec: DLA}, we describe the model architecture, parameters, and datasets employed in the DLA. Section \ref{sec: R&D} presents the experimental results and discussion, including performance analysis and comparative evaluation of two algorithms. Finally, Section \ref{sec: summary} concludes the paper and outlines potential avenues for future work.

\section{The TAO detector}
\label{sec:TAO detector}
\begin{figure}
     \centering
     \includegraphics[width=0.72\linewidth]{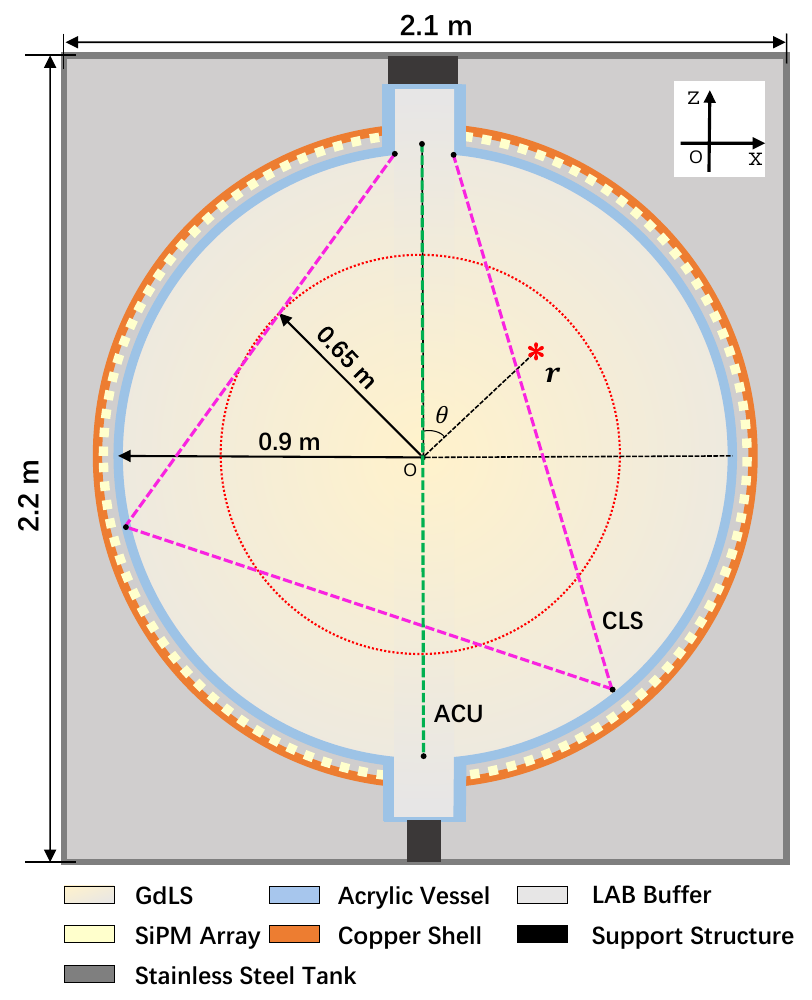}
     \caption{Structural Schematic of the TAO's central detector. The X-axis and Z-axis of the detector are shown in the upper right corner, with the origin at the center of the spherical GdLS volume. The event vertex ($\ast$) is expressed in spherical coordinates as $\bm{r}=(R, \theta, \phi )$, where $R$ is the vertex radius, $\theta$ the polar angle, and $\phi$ the azimuthal angle. The green dashed line represents the ACU calibration path along the Z-axis, and the pink dashed line indicates the CLS calibration path.}
     \label{fig: CD}
 \end{figure}

The CD configuration of the TAO experiment is depicted in Fig.~\ref{fig: CD}.
The innermost component of the CD is the liquid scintillator system, which consists of a spherical acrylic vessel (AV) with an inner diameter of 1.8~$\mathrm{m}$, filled with 2.8 tonnes of GdLS. To mitigate external radioactive and incomplete energy deposition near the GdLS boundary, a fiducial volume of 0.65~$\mathrm{m}$ radius is designated as the event selection criterion. 

A total of 4024 Hamamatsu S16088 Silicon Photomultiplier (SiPM) tiles, each measuring $50.7 \times 50.7~\mathrm{mm}^2$, are installed in an array configuration on the surface of an acrylic vessel. The acrylic vessel has a thickness of $2~\mathrm{cm}$ and is supported by a copper shell structure.
Each tile consists of 32 SiPMs with active areas of $12 \times 6~\mathrm{mm}^2$, and includes two readout channels, resulting in a total of 8048 readout channels. The SiPM tile coverage in the TAO central detector surpasses 93.5\%. This high coverage substantially improves photon collection efficiency, thereby ensuring an energy resolution $\sim$ 2\% at 1 MeV. The data acquisition system is built upon the AD9083 ADC chip for data readout~\cite{Venettacci:2024hwy}, providing a 16-bit resolution and a sampling rate of $\rm125~MS/s$. This corresponds to a time resolution of $\rm 8~ns$. In a compact detector like TAO (with a radius $\rm R < 0.9~m$), this level of time resolution becomes the primary factor limiting timing measurement precision, thereby restricting the utility of timing information in the event reconstruction process.

At an operating temperature of $-50\,^{\circ}\text{C}$, the dark noise (DN) rate of the SiPM is less than $\mathrm{100~ \mathrm{Hz/mm}^2}$.
The copper shell is housed in a stainless steel tank measuring 2.1 m in diameter and 2.2 m in height, which is filled with linear alkylbenzene (LAB) as a buffer liquid to shield external radiation, maintain temperature stability, and enable optical coupling between the acrylic container and the SiPM surface. 
In addition, the TAO detector features two calibration systems, namely an Automatic Calibration Unit (ACU) and a Cable Loop System (CLS). These systems perform regular calibrations using LED light sources and radioactive sources to correct the energy non-uniformity \cite{Xu:2022mdi}. The detector also incorporates a multi-layer shielding system consisting of a water shield, a high-density polyethylene (HDPE) layer, and a lead layer, to reduce environmental radiation. Additionally, a plastic scintillator veto system is employed to suppress cosmic ray backgrounds~\cite{Luo:2023inu}.

The TAO detector cannot directly provide the vertex $\bm{r}$ or energy information of detected particles, necessitating event reconstruction based on charge and timing data from SiPMs. An efficient reconstruction algorithm will significantly improve event selection efficiency, accelerating the experiment's physical objectives. The TAO detector is currently being installed and deployed at the Taishan Nuclear Power Plant. Monte Carlo simulations were employed within the TAO offline software to prepare for data processing. These simulations encompassed comprehensive modeling of the detector, electronics, and calibration, enabling a more realistic characterization of the detector’s charge and timing responses \cite{JUNO:2020ijm}, which served as the data source for the development and evaluation of the reconstruction algorithms.

\section{The charge center algorithm}
\label{sec: CCA}
\subsection{Principles}

\begin{figure}
    \centering
    \includegraphics[width=0.4\linewidth]{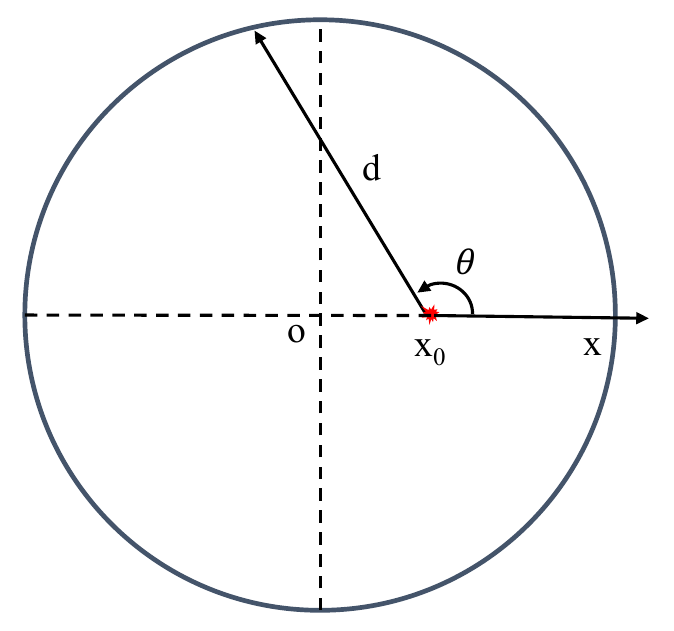}
    \caption{Geometric schematic of photon propagation in a spherical detector. It is assumed that the photons emitted at position $x_0$ propagate in straight lines and are completely detected on the spherical surface.}
    \label{fig: α}
\end{figure}

The CCA is a widely used vertex reconstruction technique in particle physics experiments \cite{DayaBay:2012aa}. The essence of this method is to determine the interaction vertex by calculating the charge-weighted average of the positions of all the photosensors, such as PMTs or SiPMs, within the detector. The reconstructed vertex based on the CCA for a spherical detector can be expressed as follows:
\begin{equation}
\vec{r}_\mathrm{cc} =  \frac{\sum_i q_i \overrightarrow{r_i}}{\sum_i q_i}=\alpha\cdot\vec{r}_\mathrm{true}, 
\label{Eq: CCA}
\end{equation}
where $\vec{r}_\mathrm{cc}$ and $\vec{r}_\mathrm{true}$ represent the reconstructed and true vertex position vectors, respectively; 
$q_i$ quantifies the charge detected by the \(i\)-th photosensor, $\overrightarrow{r_i}$ corresponds to the positional vector of said photosensor; 
$\sum_i q_i$ represents the summation of all charges acquired by the photosensor array; the parameter $\alpha$ serves as a scaling factor under ideal conditions where photons emitted from the point-like source propagate without attenuation through the medium and undergo complete absorption by the surface-mounted photosensors.
The $\alpha$ depends only on the detector geometry and can be calculated via simple assumptions. 
As depicted in Fig. \ref{fig: α}, 
It is assumed that a point light source is located at $x_0$ on the X-axis of a unit sphere (radius = 1).
Taking into account the geometric symmetry of the spherical detector, the reconstructed vertex can be calculated using the continuous-limit form of Eq. (\ref{Eq: CCA}):
\begin{equation}x_{\mathrm{cc}}=\frac{1}{4\pi}\int x(\theta,~\phi) d\Omega=\frac{1}{4\pi}\int_0^{2\pi}d\phi\int_0^\pi(x_0+d(x_0,~\theta)\cdot\cos\theta)\sin\theta d\theta=\frac{2}{3}x_0,
\label{Eq:α caculation}
\end{equation}
where $x_\mathrm{cc}$ is the reconstructed point light source position along the X-axis, with coordinates zero on the Y- and Z-axes, and $d(x_0,~\theta)$ denotes the distance from $x_0$ to the light collection position. 
By performing a comparative analysis of Eq. (\ref{Eq: CCA}) and Eq. (\ref{Eq:α caculation}), the scaling factor for a spherical detector is mathematically derived as $\alpha = 2/3$.

Although the inherent geometric effects can be readily corrected for spherical detectors, real-world scenarios often introduce a host of non-ideal factors. In the case of the TAO detector, these include:
\begin{enumerate}
    \item As shown in Fig. \ref{fig: CD}, the SiPMs cannot fully cover the spherical surface, with two $\mathrm{\sim25~cm}$ diameter openings along the Z-axis lacking photosensors (hereafter referred to as the "dual-opening");
    \item Reflections on the SiPM surface directly impact its photon detection efficiency;
    \item DN, cross-talk (CT), afterpulsing (AP), readout channel saturation, and other electronic effects directly impact SiPM charge counting~\cite{Guan:2023uaa,Zhao:2022gks}; specifically, the DN count rate of $\mathrm{20~ {Hz/mm}^2}$, CT probability of 16\%, and channel saturation threshold of 2 V are set in the simulation program based on preliminary results from batch testing;
    \item Photon attenuation due to absorption in the GdLS.
\end{enumerate}
Therefore, to mitigate these effects and improve the reconstruction accuracy of CCA, ensuring its reliability for future applications in real data reconstruction, we need to implement some necessary optimizations.

\subsection{Optimizations}
\label{sec:improvements}
The data used to improve CCA comes from Monte Carlo simulations performed along the calibration paths of CLS and ACU, as shown in Fig.~\ref{fig: CD}. During experimental phases, these data can be obtained through the calibration system, thereby ensuring the reliability and applicability of the optimization for the CCA. Among the various sources in the calibration system, the $^{137}$Cs source emits gamma rays with a specific energy of 0.662 MeV, making it particularly suitable for small-scale detectors like TAO and ideal for correcting the CCA. Given the extensive coverage of SiPMs in TAO and the inherent symmetry of the spherical detector, the original CCA already attains adequate reconstruction accuracy for the angles $\theta$ and $\phi$ of the vertex.  The specific reconstruction performance can be referenced in Section \ref{subsec:comparion2algorithms}. Therefore, we mainly optimize the reconstruction of the vertex radius $R$ here.

\begin{figure}
    \centering
    \begin{subfigure}[b]{0.49\textwidth}
        \includegraphics[width=1\textwidth]{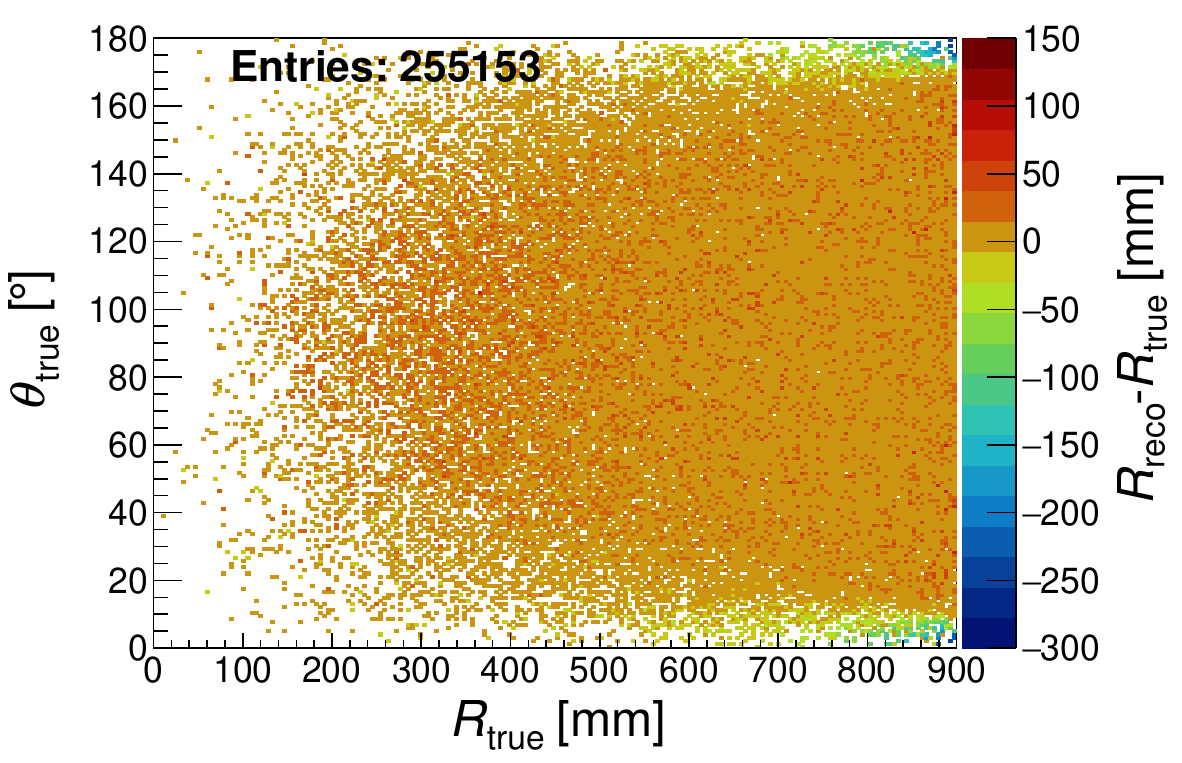}
    \caption{}
    \label{fig: RBias2a}
    \end{subfigure}
    \begin{subfigure}[b]{0.49\textwidth}
        \includegraphics[width=1\textwidth]{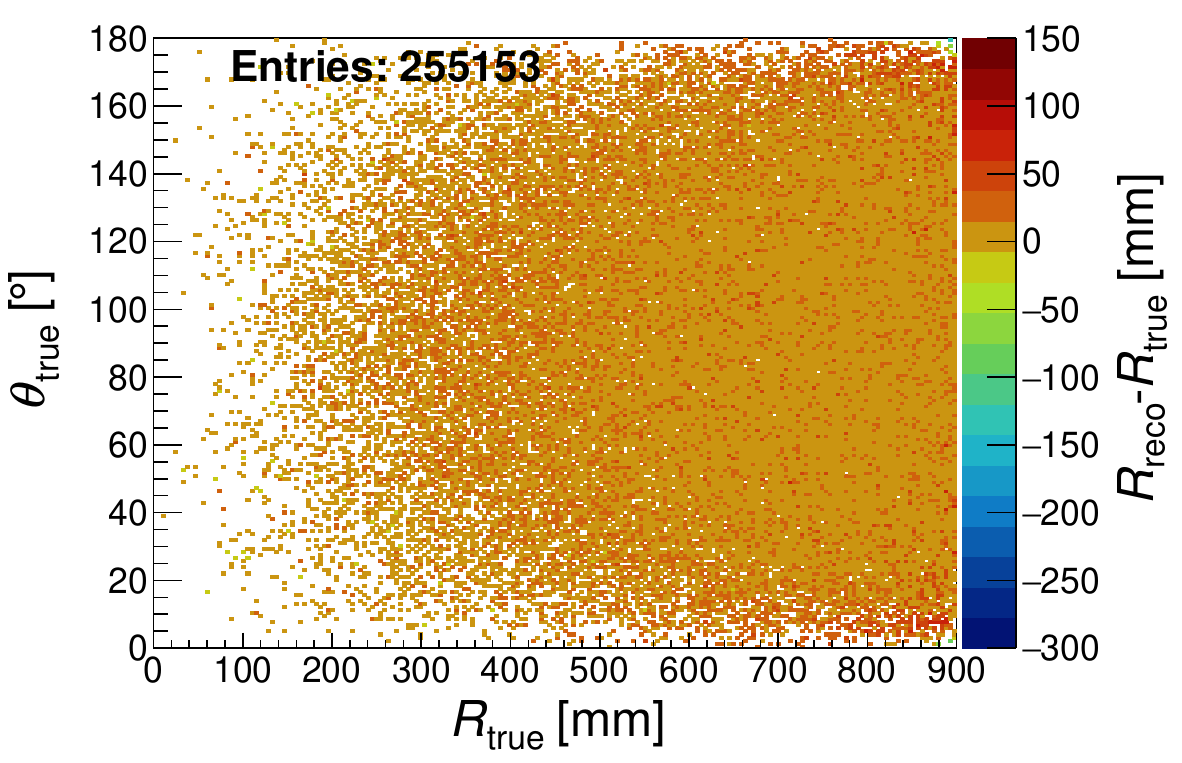}
        \caption{}
        \label{fig: RBias2b}
    \end{subfigure}
    \caption{Impact of TAO detector's intrinsic openings on the $R$ reconstruction bias, defined as the difference between $R_\text{reco}$ and $R_\text{true}$. Panels (a) and (b) respectively show the results before and after applying the dual-opening correction. }
    \label{fig: RBias}
\end{figure}

\begin{figure}
    \centering
    \begin{subfigure}[b]{0.49\textwidth}
        \includegraphics[width=0.9\textwidth]{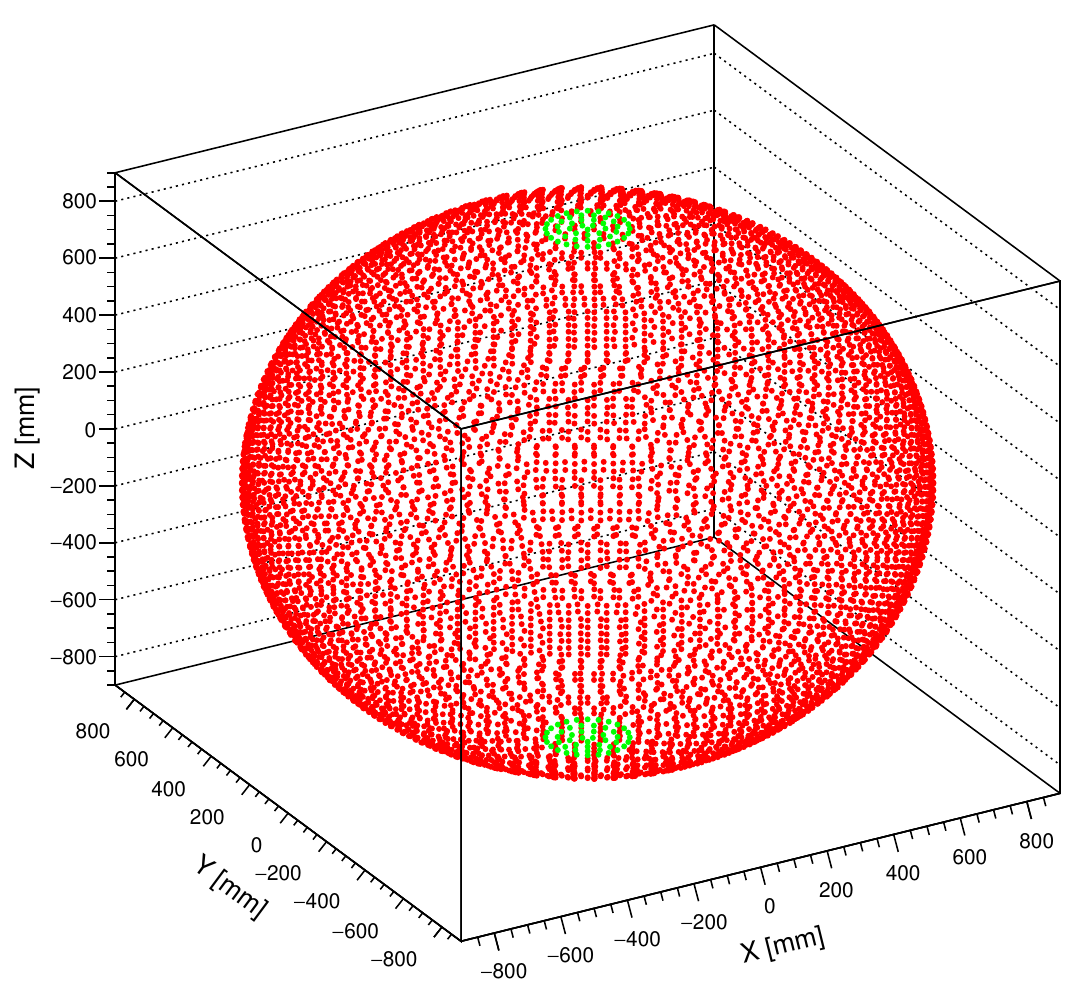}
    \caption{}
    \label{fig: virtualPoint2a}
    \end{subfigure}
    \begin{subfigure}[b]{0.49\textwidth}
        \includegraphics[width=0.9\textwidth]{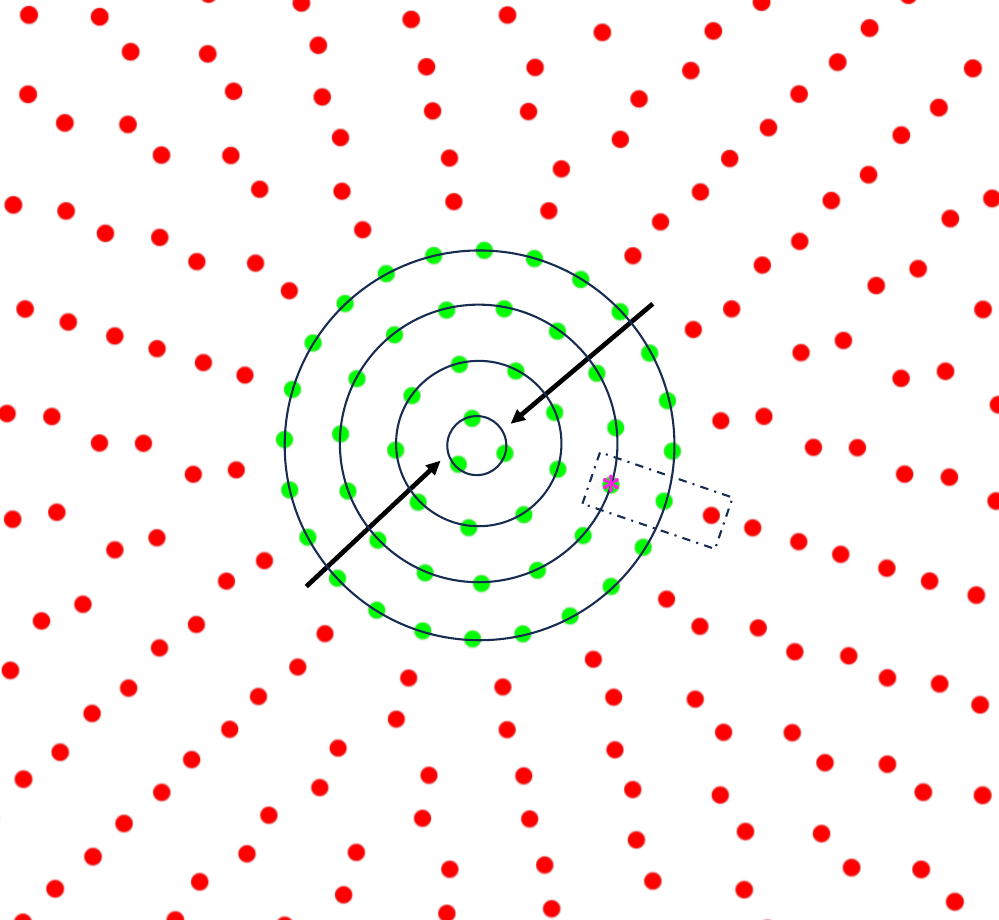}
        \caption{}
        \label{fig: virtualPoint2b}
    \end{subfigure}
    \caption{Schematic diagram of virtual channel construction. Each SiPM has two channels, and the TAO detector has a total of 8048 channels. (a) Distribution of virtual channels in the detector. (b) Charge calculation for virtual channels, with arrows indicating the calculation direction and the dashed box illustrating the interpolation process.}
    \label{fig: virtualPoint}
\end{figure}

First, we consider a simple case that utilizes Eq. (\ref{Eq: CCA}) to calculate the $R$ reconstruction results for electrons directly without electronic effects. Figure \ref{fig: RBias} presents the reconstruction results for electrons with energies ranging from 1 MeV to 9 MeV, uniformly generated within the GdLS. It is observed that the reconstruction bias $B_\text{reco}$ becomes more pronounced near the dual-opening location, where $B_\text{reco}=R_\text{reco}-R_\text{true}$, with $R_\text{reco}$ and $R_\text{true}$ representing the reconstructed radius and the true radius of the energy deposition position, respectively. This is primarily due to an increase in photon leakage close to the dual-opening, resulting in a lower $R_\text{reco}$. To address the impact of the dual-opening effect, we propose the following dual-opening correction scheme:

\begin{enumerate}
    \item Create 52 uniformly distributed virtual channels at the upper and lower openings, as shown in Fig. \ref{fig: virtualPoint2a};
    \item Assume that the number of photons received by the virtual channels is equal to that received by the channels symmetric to the energy deposition point, and their charge numbers can be approximated as equal;
    \item Calculate the charge of the virtual channels point by point from the outer to the inner;
    \item For points where symmetric channels remain virtual, use interpolation from neighboring points, as illustrated in Fig. \ref{fig: virtualPoint2b}.
\end{enumerate}

\begin{figure}
    \centering
    \includegraphics[width=0.8\linewidth]{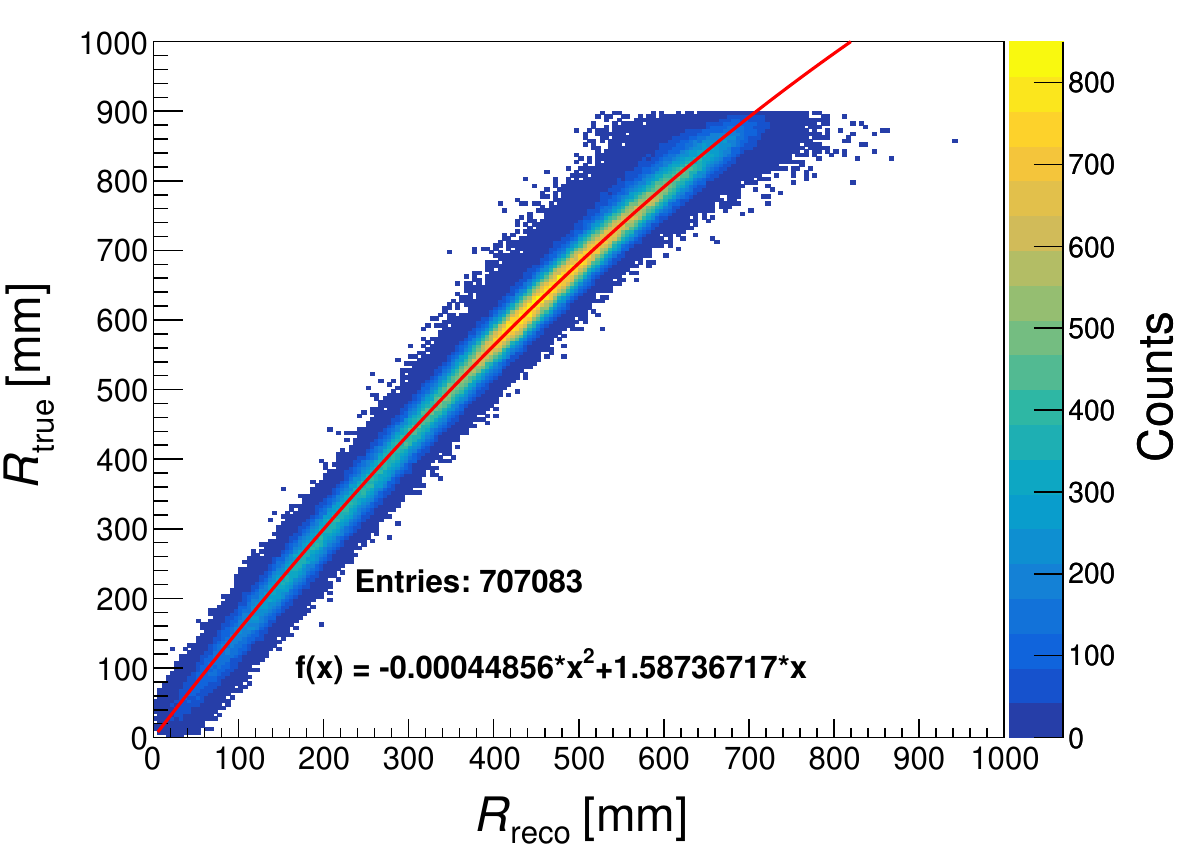}
    \caption{After calibration with $^{137}$Cs, a two-dimensional distribution of $\mathit{R_\text{true}}$ versus  $\mathit{R_\text{reco}}$ was obtained and fitted with a quadratic function (red curve). Dual-opening correction was applied, while the DN remains unprocessed. }
    \label{fig: Cs-137_Rcc_Rtrue}
\end{figure}

\begin{figure}
    \centering
    \begin{subfigure}[b]{0.49\textwidth}
        \includegraphics[width=1\textwidth]{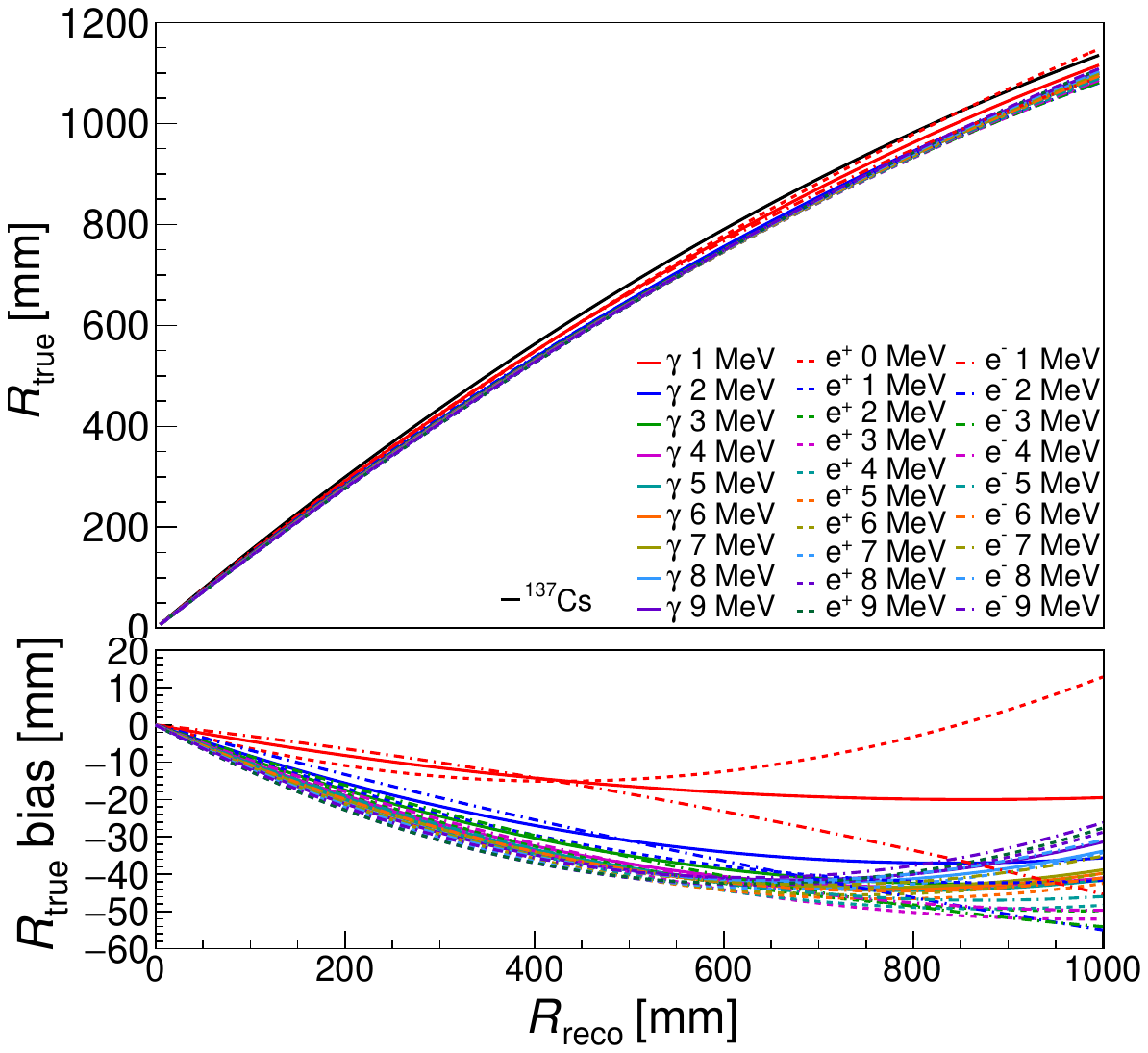}
    \caption{}
    \label{fig: fitCurveA}
    \end{subfigure}
    \begin{subfigure}[b]{0.49\textwidth}
        \includegraphics[width=1\textwidth]{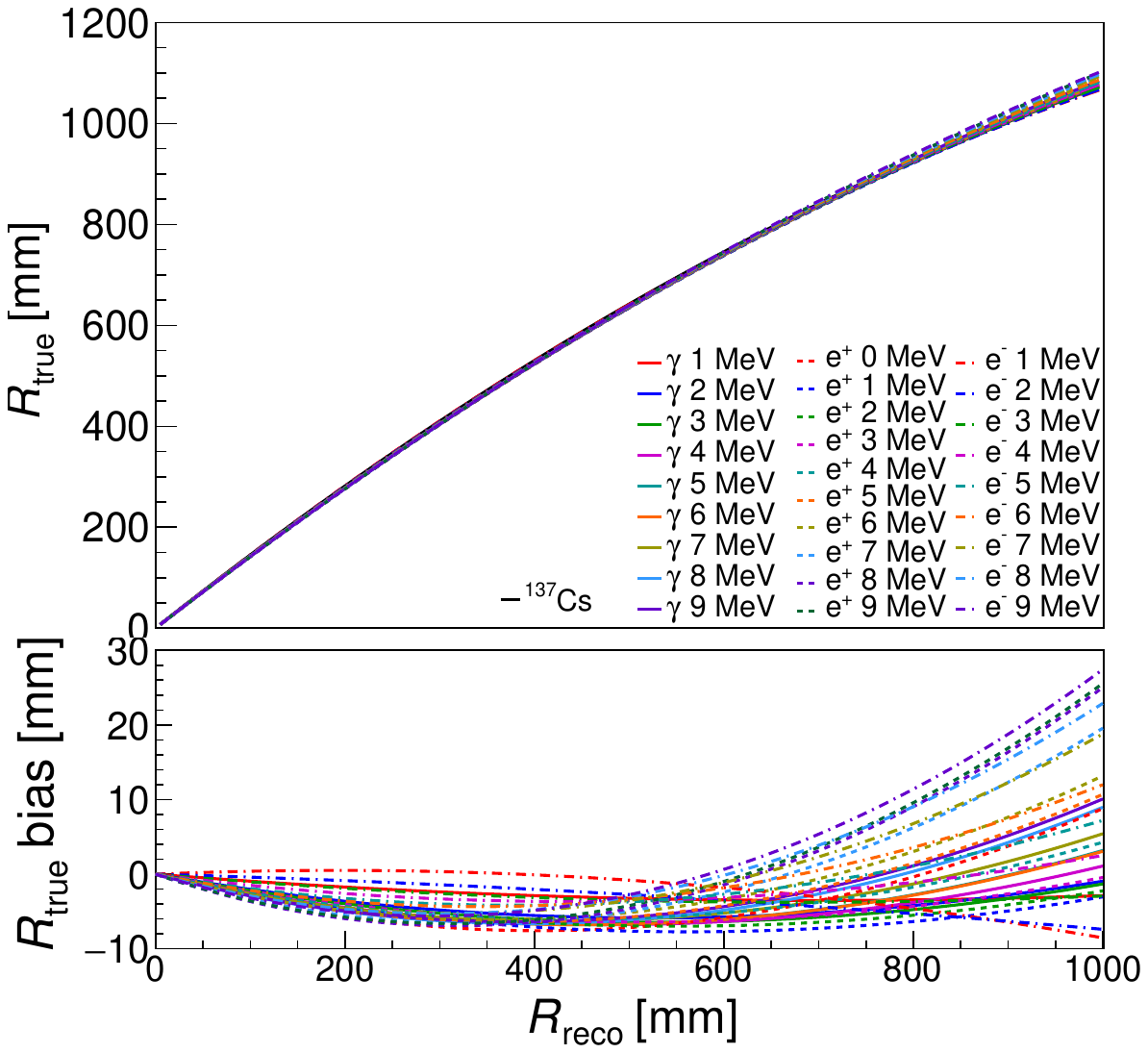}
        \caption{}
        \label{fig: fitCurveB}
    \end{subfigure}
    \caption{Fitting curves of the two-dimensional distributions of $\mathit{R_\text{true}}$ and $\mathit{R_\text{reco}}$. The $ ^{137} $Cs curve (E = 0.662 MeV) is based on simulations along the calibration path, while the gamma ($\gamma$), positron ($e^+$), and electron ($e^-$) curves are derived from uniform simulations in the GdLS. Panels (a) and (b) show the results without and with DN correction, respectively. ``$R_{\rm true}$ bias" represents the difference in the $R_{\rm true}$ values between other curves and the $ ^{137} $Cs curve.}
    \label{fig: fitCurves}
\end{figure}
\begin{figure}[h]
    \centering
    \includegraphics[width=0.8\linewidth]{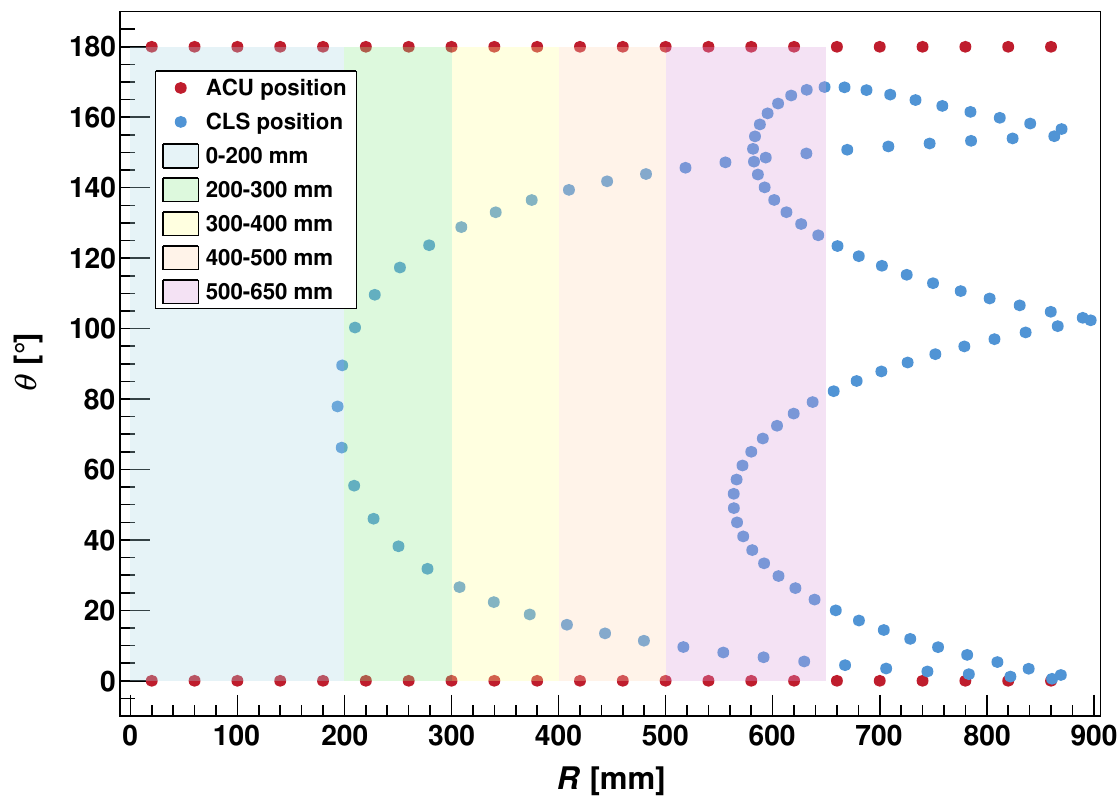}
    \caption{Spatial distribution of calibration points. Multi-correction curve fitting scheme: the fiducial volume is partitioned into five regions, with calibration data in each interval independently fitted to derive five distinct curves; for areas outside the fiducial volume, a single correction curve fitting scheme is applied.}
    \label{fig:CalibPoints}
\end{figure}

After applying the dual-opening correction scheme, significant improvements in reconstruction bias were observed (see Fig. \ref{fig: RBias2b}), with the negative distortions near the openings substantially reduced.
Subsequently, we incorporated a complete electronics and calibration simulation process to reflect real data better. By calibrating with a $^{137}$Cs source, we established a map of $\mathit{R_\text{reco}}$ versus $\mathit{R_\text{true}}$, as depicted in  Fig. \ref{fig: Cs-137_Rcc_Rtrue}. A second-order polynomial model was fitted to derive the nonlinear relationship between $\mathit{R_\text{true}}$ and $\mathit{R_\text{reco}}$, given by: 
\begin{equation}
\mathit{R_\text{true}} = a\cdot\mathit{R_\text{reco}^\mathrm{2}}+b\cdot\mathit{R_\text{reco}}, 
\label{Eq: CCAFitCurve}
\end{equation}
where $a$ and $b$ are fitting coefficients. This relationship enables the correction of the original result $\mathit{R_\text{reco}}$ to obtain a corrected reconstructed radius $\mathit{{R}^\text{corr}_\text{reco}}$~$(\simeq \mathit{{R}_\text{true}})$. However, electronic effects, mainly DN, prevent this model from correcting particles effectively across the entire energy range. This limitation arises because the DN counts of each SiPM tile are sampled from a Poisson distribution with a fixed DN rate in the TAO offline software, introducing random fluctuations that degrade correction accuracy. As shown in Fig.~\ref{fig: fitCurveA}, without DN correction, the $R$ reconstruction fitting curves for particles of different energies exhibit inconsistency, with divergence observed at larger vertex radius values. This discrepancy arises because the expected number of DN per event remains relatively stable, causing the proportion of DN counts to decrease as the particle energy increases. Consequently, DN has a more pronounced impact on event reconstruction for low-energy particles, necessitating a correction to ensure accurate results. In the presence of DN, Eq. (\ref{Eq: CCA}) can be expressed as follows:
\begin{equation}\begin{aligned}
 & \vec{r}_\mathrm{cc} =\frac{\sum_{i}q_i^{p}\cdot\vec{r}_\mathrm{cc}^p+\sum_iq_i^{dn}\cdot\vec{r}_\mathrm{cc}^{dn}}{\sum_iq_i}=\frac{(\sum_iq_i-\sum_{i}q_i^{dn})\cdot\vec{r}_\mathrm{cc}^p}{\sum_iq_i}, 
\end{aligned}\end{equation}
where $\sum_{i}q_i^{p}$ and $\sum_iq_i^{dn}$ denote the total charges induced by photons and DN, respectively; The vertex positions calculated from these charges are $\vec{r}_\mathrm{cc}^p$ (photon-induced) and $\vec{r}_\mathrm{cc}^{dn}$ (DN-induced), and $\vec{r}_\mathrm{cc}^{dn} \approx 0$ resulting from the symmetric layout of the SiPM array. As illustrated in Fig. \ref{fig: fitCurveB}, the results show a significant improvement in the consistency of fitting curves for particles of different energies after DN correction.

It should be noted that, during the construction of the $^{137}$Cs fitting curve, calibration points were systematically selected at 40 mm intervals along both CLS and ACU calibration paths. The spatial distribution of these calibration points is presented in Fig.~\ref{fig:CalibPoints}. If a single correction curve (SCC) fitting is directly applied to all data points, as shown in Fig.~\ref{fig: Cs-137_Rcc_Rtrue}, the resulting curve would exhibit a bias toward regions with denser point distributions while deviating from sparser regions, due to the non-uniform radial distribution of the calibration points. To address the reconstruction errors introduced by this fitting approach, we developed a multi-correction curve (MCC) fitting scheme, as illustrated in Fig.~\ref{fig:CalibPoints}. A detailed analysis of the reconstruction performance of these two fitting schemes is provided in Section~\ref{subsec:impacts4CCA}.

\section{The deep learning algorithm}
\label{sec: DLA}

Deep learning is a class of machine learning algorithms that utilizes multiple layers of information abstraction to capture the intrinsic representation of complex data. Over the past decade, it has experienced significant advancements in particle physics, driven by the evolution of computational technology. Its robust data processing capabilities have demonstrated remarkable potential in addressing key challenges, including event reconstruction \cite{Li:2022gpb, Qian:2021vnh, Li:2022tvg}, signal-background discrimination \cite{Baldi:2014kfa,Chan:2019fuz}, spectrum decomposition \cite{Zeng:2023att,Chen:2024vdq}, and particle identification \cite{Psihas:2019ksa,Lee:2022kdn}. In the TAO experiment, the nonlinear modeling capabilities of deep learning, rooted in its multi-layer neural network architecture, can effectively resolve detector response features that are difficult to capture with traditional algorithms, thereby significantly enhancing reconstruction accuracy. Furthermore, its parallel computing framework is anticipated to greatly improve the efficiency of vertex reconstruction tasks when processing vast amounts of real data.

\subsection{Dataset}
We constructed three independent datasets, namely training, validation, and test datasets, to investigate the performance of deep learning algorithms in vertex reconstruction for the TAO experiment. The training dataset is used for model fitting, the validation dataset for tuning and monitoring performance during training, and the test dataset is reserved for the final evaluation of the trained model. The training and validation datasets contain 240,000 and 60,000 events, respectively, evenly divided among positron, electron, and gamma types. For both datasets, the initial kinetic energies of the particles are uniformly distributed between 0 and 10 MeV, and the initial positions are uniformly sampled within the GdLS region of the TAO's CD. The test dataset is generated using a similar procedure, with initial kinetic energies at 1 MeV intervals: 0 to 9 MeV for positrons, and 1 to 9 MeV for electrons and gamma rays. For each kinetic energy, 20,000 events are generated and uniformly distributed within the GdLS. Notably, this test dataset is identical to that used in CCA, thus ensuring a fair and consistent comparison between the two algorithms.

\begin{figure}
    \centering
    \begin{subfigure}[b]{0.49\textwidth}
        \includegraphics[width=1\textwidth]{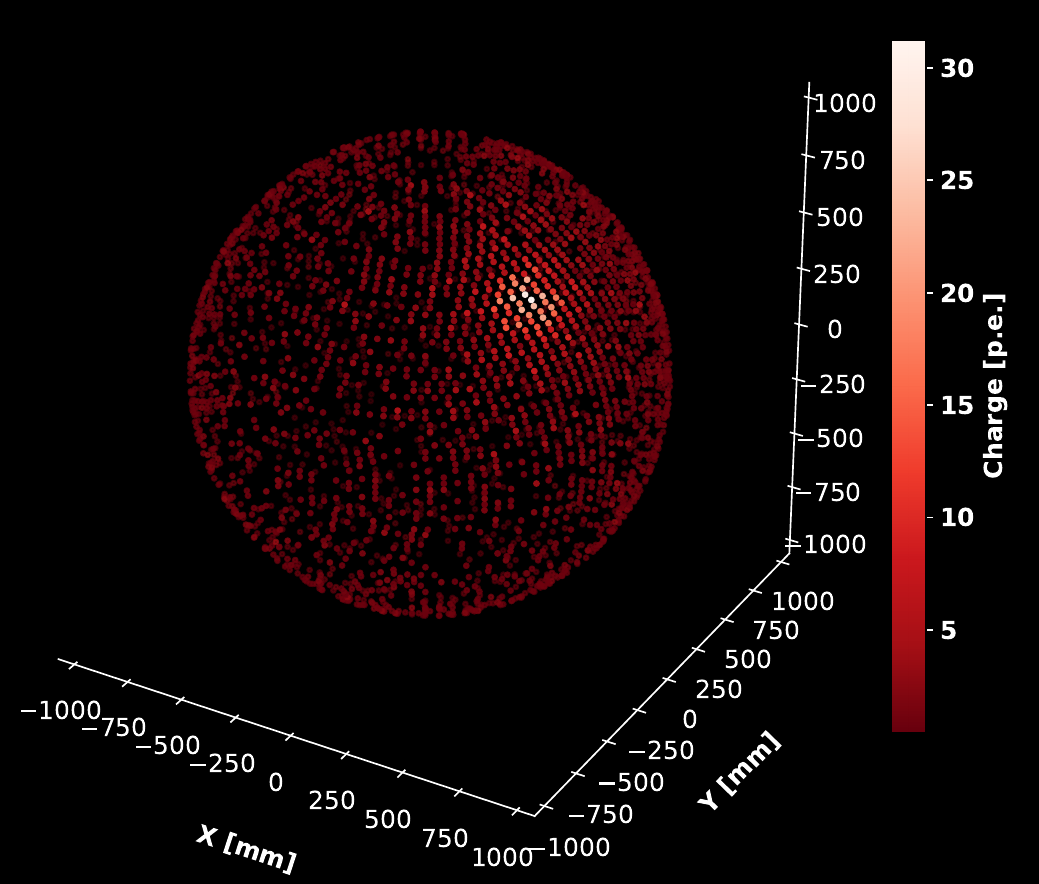}
    \caption{}
    \label{fig: Charge3D}
    \end{subfigure}
    \begin{subfigure}[b]{0.49\textwidth}
        \includegraphics[width=1\textwidth]{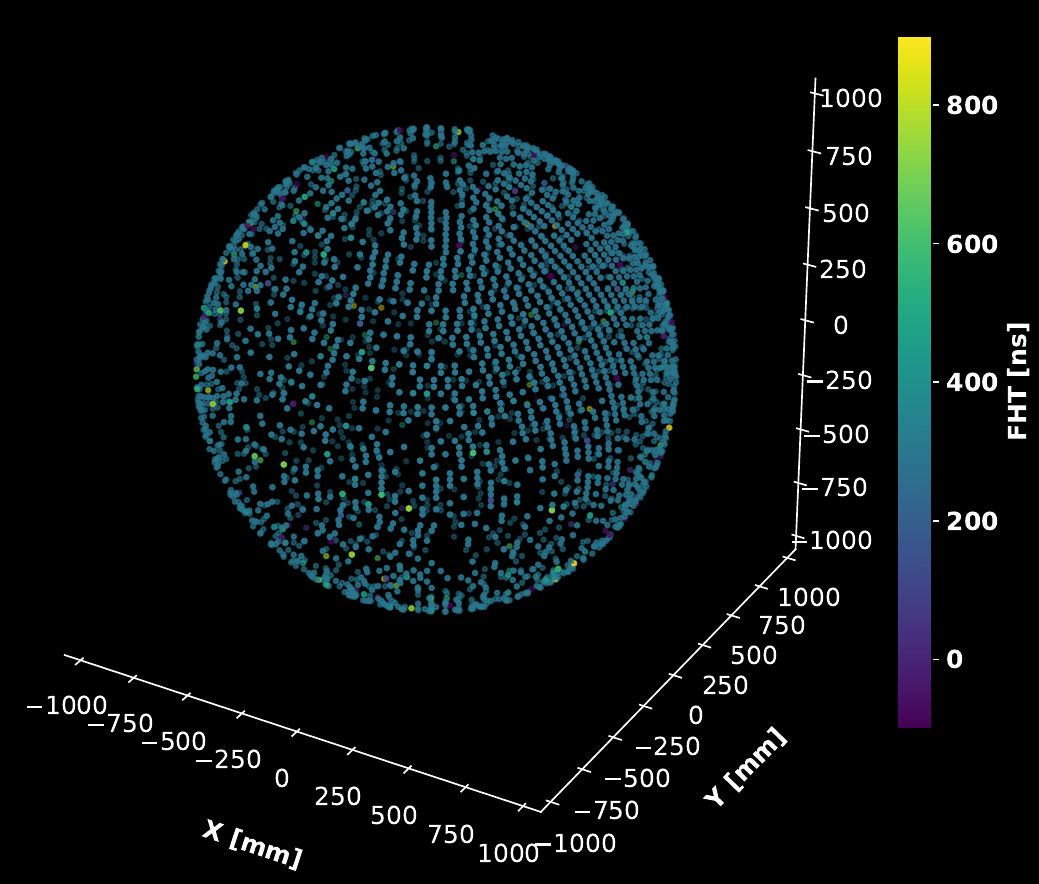}
        \caption{}
        \label{fig: FHT3D}
    \end{subfigure}
    \caption{The response characteristics of the SiPM array to 1 MeV electron events: the (a) and (b) panels show the charge response and the first hit time response, respectively. Channels without hits are shown in black.}
    \label{fig: 3D}
\end{figure}

The model's input consists of two components: the charge distribution and the across-channel first hit time (FHT) distribution, as illustrated in Fig.~\ref{fig: 3D}. The three-dimensional (3D) arrangement of SiPMs in TAO presents unique challenges for applying CNNs commonly used in deep learning. To fully leverage the capabilities of CNNs, it is necessary to transform the information detected by the SiPMs into two-dimensional (2D) images compatible with model processing. Specifically, this transformation should ensure non-overlapping channel data while minimizing spatial distortion, thereby preserving an accurate representation of the spatial characteristics of each event. We have implemented a lossless projection method to generate the 2D images, as shown in Fig.~\ref{fig: 2D}. For the $i$-th readout channel, the pixel coordinates in the $\theta$ and $\phi$ directions, $\mathit{\theta^\text{index}_\text{i}}$ and $\mathit{\phi^\text{index}_\text{i}}$, are determined by the following formulas:

\begin{figure}
    \centering
    \begin{subfigure}[b]{0.85\textwidth}
        \includegraphics[width=1\textwidth]{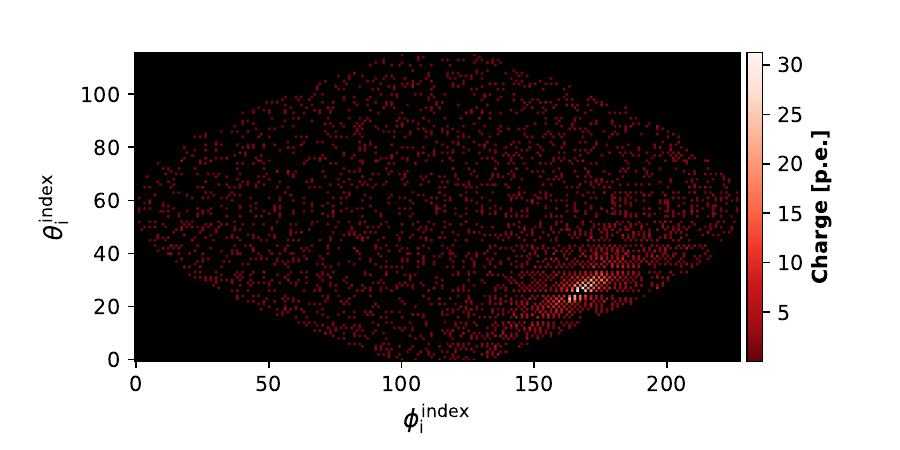}
        \caption{}
        \label{fig: Charge2D}
    \end{subfigure}
    \begin{subfigure}[b]{0.85\textwidth}
        \includegraphics[width=1\textwidth]{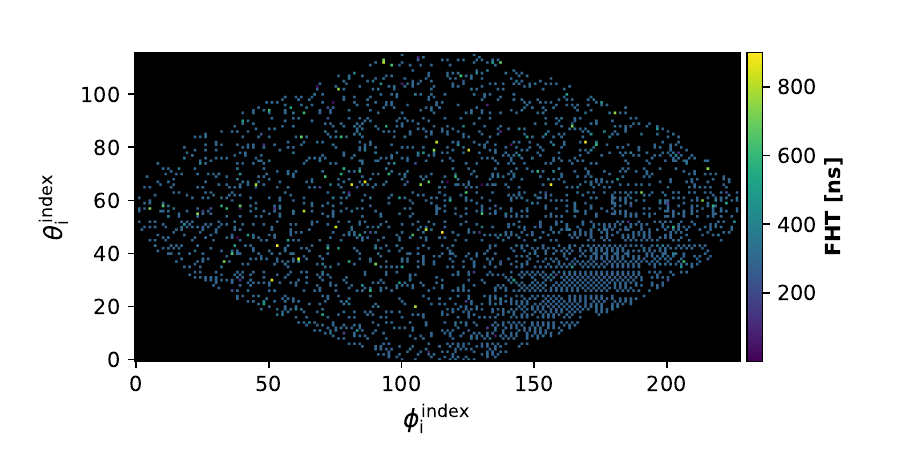}
        \caption{}
        \label{fig: FHT2D}
    \end{subfigure}
    \caption{The 2D projection of 3D features: the top and bottom panels are the projection of the charge and FHT responses, respectively. }
    \label{fig: 2D}
\end{figure}

\begin{equation}
\begin{aligned}
    \mathit{\theta^{\text{index}}_i} &= \text{floor}\left((\theta_i-\theta_\text{min}) \times \frac{H_{\text{max}}}{\pi}\right),  \\
    \mathit{\phi^{\text{index}}_i} &= \left[N_\mathrm{max}\cdot\frac{\sqrt{R^2-z^2}}{R}\cdot\frac{2\cdot\arctan(x/y)}{\pi}\right] + {N_\mathrm{max}}, 
\end{aligned}
\end{equation}
where $\theta_i$ is the zenith angle of the $i$-th readout channel, $\text{H}_{\text{max}}$ is the max pixel height with a value of 128, and $\text{“floor”}$ signifies the floor operation; $x$, $y$, and $z$ represent the global coordinates of the readout channel, $R$ is the radius at which the readout channel is positioned, and $\text{N}_{\text{max}}$ is the maximum number of channels per row, set to 114. The calculated optimal image size is $116\times228$.

In addition to shape transformation, we normalize the features using the following formula: 

\begin{equation}
    z = \frac{x - \mu}{\sigma}, 
    \label{Eq:Z-Score}
\end{equation}

where $\mu$ is the mean and $\sigma$ is the standard deviation of the feature. This normalization process ensures that the features have zero mean and unit variance, which mitigates the impact of differing scales among input features, accelerates the training process, and improves the stability and convergence of the model.

\subsection{The VGG-T model}

The Visual Geometry Group (VGG) at the University of Oxford proposed a classic CNN model known as the VGG network architecture in 2014 \cite{Simonyan:2014cmh}. This model is characterized by its deep architecture, which allows for extracting hierarchical features from input data. Its advantages include improved performance on various image classification tasks, a straightforward design that facilitates transfer learning, and robustness to different image resolutions. These features make the VGG network popular for many computer vision applications.  In the field of physics, the VGG network has also been widely applied. For instance, they have demonstrated promising results in pp solar neutrino identification~\cite{Chen:2023xhj}, dynamic aperture prediction in accelerator experiments~\cite{DiCroce:2024dfy} , as well as event reconstruction tasks in neutrino experiments such as PandaX~\cite{Li:2022gpb}, CHIPS~\cite{Tingey:2022evd}, and JUNO~\cite{Qian:2021vnh}. In this work, building upon the VGG framework, we designed a tailored VGG-T model to predict the vertex of TAO events.

\begin{figure}
    \centering
    \includegraphics[width=1\linewidth]{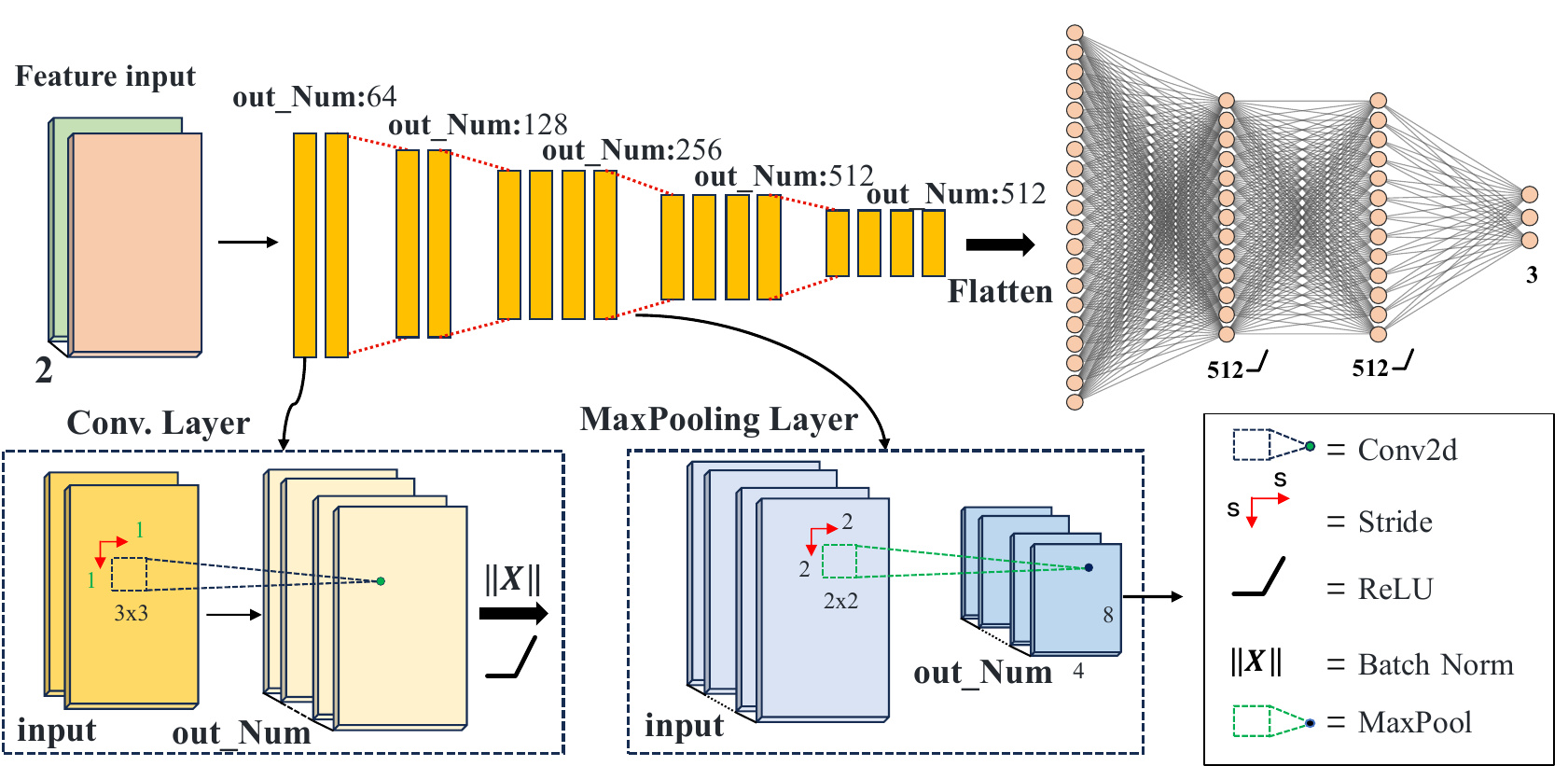}
    \caption{Network structure of VGG-T. Each convolution layer uses a $3 \times 3$ kernel and a stride of 1. Each max-pooling layer use a $2 \times 2$ pooling kernel and a stride of 2.}
    \label{fig: VGGflowchart}
\end{figure}

 As illustrated in Fig.~\ref{fig: VGGflowchart}, the VGG-T model consists of sixteen convolutional layers for deep extraction of charge and time information and four max-pooling layers to reduce feature dimensions. Each convolutional layer is followed by a normalization layer to accelerate model convergence. After convolutional feature extraction, four fully-connected layers are employed to integrate features and generate vertex prediction values. The training hyperparameters are outlined in Table \ref{tab:hyperparameters}, and the training employs the Smooth L1 loss function, which is chosen for its robustness to outliers, enabling effective mitigation of gradient explosion while maintaining prediction accuracy \cite{Girshick:2015ICCV}. 
 
 The predicted vertex coordinates are represented in the Cartesian space as $\mathbf{r}(x, y, z)$, with values constrained to the range [-900 mm, 900 mm]. Adopting Cartesian coordinates provides a continuous and unambiguous representation, avoiding the instabilities caused by the periodicity of angular parameters in spherical coordinates and thus simplifying the design of the loss function. To enforce this constraint, the loss function incorporates two penalty terms: $\text{Penalty}_{\text{upper}}$ for values exceeding 900 mm and $\text{Penalty}_{\text{lower}}$ for values below -900 mm. The total loss function is defined as follows:
\begin{align}
    \mathcal{L}_{\text{total}} &= \mathcal{L}_{\text{smooth}_{L1}} + \lambda \left( \text{Penalty}_{\text{upper}} + \text{Penalty}_{\text{lower}} \right) \nonumber \\
    &= \frac{1}{n}\sum_{i=1}^n\mathrm{smooth}_{L1}(y_i-\hat{y}_i) \\
    &+ \lambda \cdot \left( \frac{1}{n}\sum_{i=1}^n\max(0,\hat{y}_i-\text{max}_{\mathrm{value}}) + \frac{1}{n}\sum_{i=1}^n\max(0,\text{min}_\mathrm{value}-\hat{y}_i) \right) \nonumber,
\end{align}
where $\lambda$ is the weight of the penalty terms used to control the strength of the penalties; $ y_i$ and $\hat{y}_i$ are the true and predicted vertex $\mathbf{r}(x, y, z)$, respectively. The definition of $\mathrm{smooth}_{L1}(y_i-\hat{y}_i)$ is as follows:  
\begin{equation}
    L_1(y_i,\hat{y_i})=
    \begin{cases}
    \frac{0.5\cdot|y_i-\hat{y}_i|^2}{\beta},\quad|y_i-\hat{y}_i|<\beta \\
     \\
    |y_i-\hat{y}_i|-0.5\cdot\beta,\quad|y_i-\hat{y}_i|\geq\beta &
    \end{cases},%~\beta = 1.
\end{equation}
where $\beta$ is a threshold to categorize prediction errors, influencing how the loss function penalizes errors, and tests show that setting $\beta$ within the range of [1, 20] has little impact on the final results. Unless otherwise specified, we adopt the default settings from Ref. \cite{Girshick:2015ICCV} in this work, with both $\lambda$ and $\beta$ set to 1.

\begin{table}[h!]
\centering
\caption{Training settings for VGG-T.}
\label{tab:hyperparameters}
\begin{tabularx}{\textwidth}{>{\bfseries}lX} 
\toprule
\textbf{Parameter} & \textbf{Value} \\
\midrule
Loss & Smooth L1 loss with penalty terms   \\
Optimizer & Adam \cite{Kingma:2014vow} ($\beta_1=0.9$, $\beta_2=0.999$) \\
Learning rate & One Cycle Learning Rate Policy \cite{smith:2018scv}, with the initial learning rate set to $1e^{-3}$. \\
Batch size & 128 \\
Number of epochs & 20 \\
\bottomrule
\end{tabularx}
\end{table}

\subsection{The ResNet-T model}

\begin{figure}
    \centering
    \includegraphics[width=0.5\linewidth]{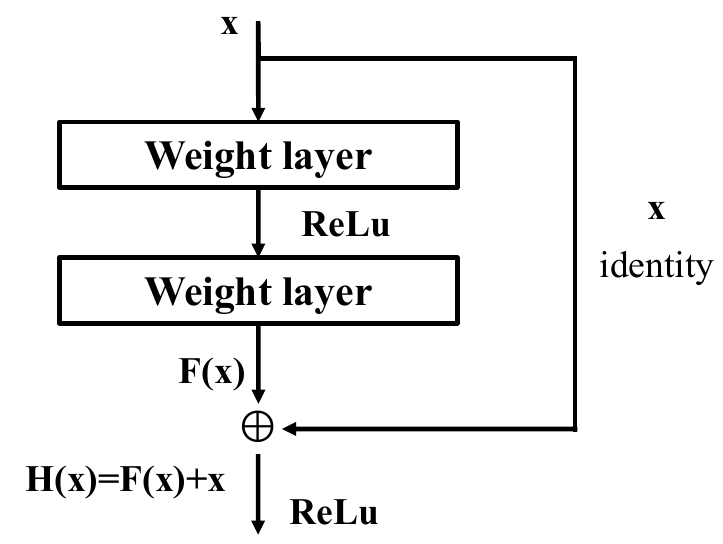}
    \caption{
    The ResNet structure uses residual connections, where $\mathbf{F(x)}$ is the output of weight layers and $\mathbf{H(x)}$ is the input to the next block. If $\mathbf{F(x)}$ is zero, then $\mathbf{H(x)}$ equals the original input $\mathbf{x}$.
    }
    \label{fig:ResidualBlock}
\end{figure}

In the VGG-T network architecture context, increasing model depth is a widely adopted strategy in deep learning to enhance the accuracy of reconstruction tasks by augmenting the model's nonlinear representation capabilities. However, deeper networks often encounter challenges such as gradient vanishing and explosion, which can hinder convergence during training. To address these issues, we implement the residual connection structure proposed in Ref. \cite{he2015:drl}, as illustrated in Fig. \ref{fig:ResidualBlock}. This structure incorporates skip connections, significantly improving feature utilization and enhancing the model's ability to learn complex patterns, facilitating a more effective optimization process.

\begin{figure}
    \centering
    \includegraphics[width=1\linewidth]{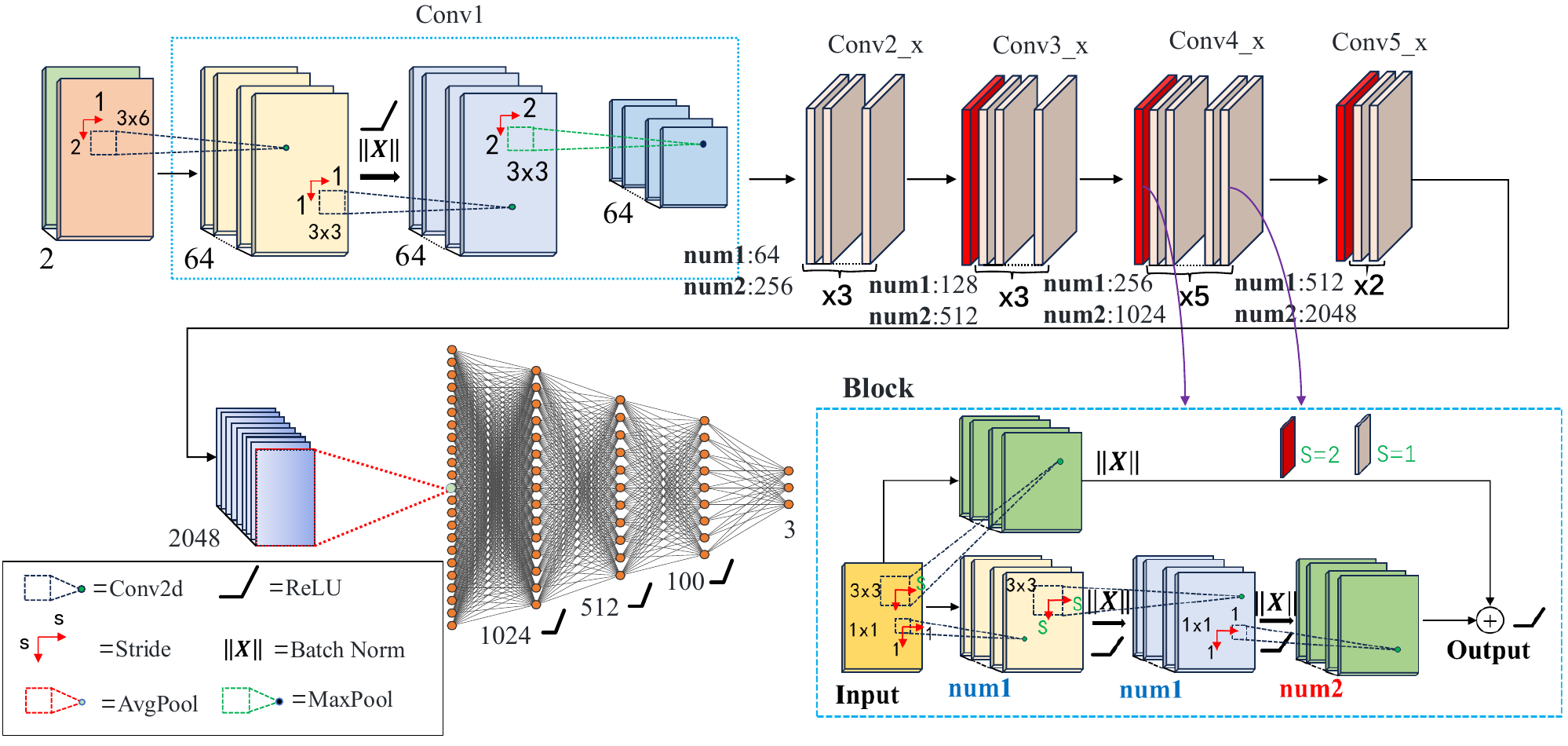}
    \caption{Network structure of the ResNet-T model. The network architecture consists of sixteen residual blocks, each containing three convolutional weight layers. After convolutional feature extraction, four fully connected layers integrate the features and output the vertex predictions.}
    \label{fig:enter-label}
\end{figure}

We utilize this technique to develop a deeper network architecture called ResNet-T, which is tailored for reconstruction tasks in the TAO experiment. As shown in Fig. \ref{fig:enter-label}, ResNet-T incorporates thirty-four convolutional layers beyond the original VGG architecture, resulting in a modest $1.6\%$ increase in the total number of parameters. Although the increase in the number of parameters is relatively small, the enhanced model capacity enables ResNet-T to capture more complex features and relationships in the data. Notably, this architectural modification leads to approximately a 36.5\% increase in GPU memory usage, while the corresponding rise in training time is about 14.4\% on an RTX-4090.

\begin{figure}
    \centering
    \includegraphics[width=0.95\linewidth]{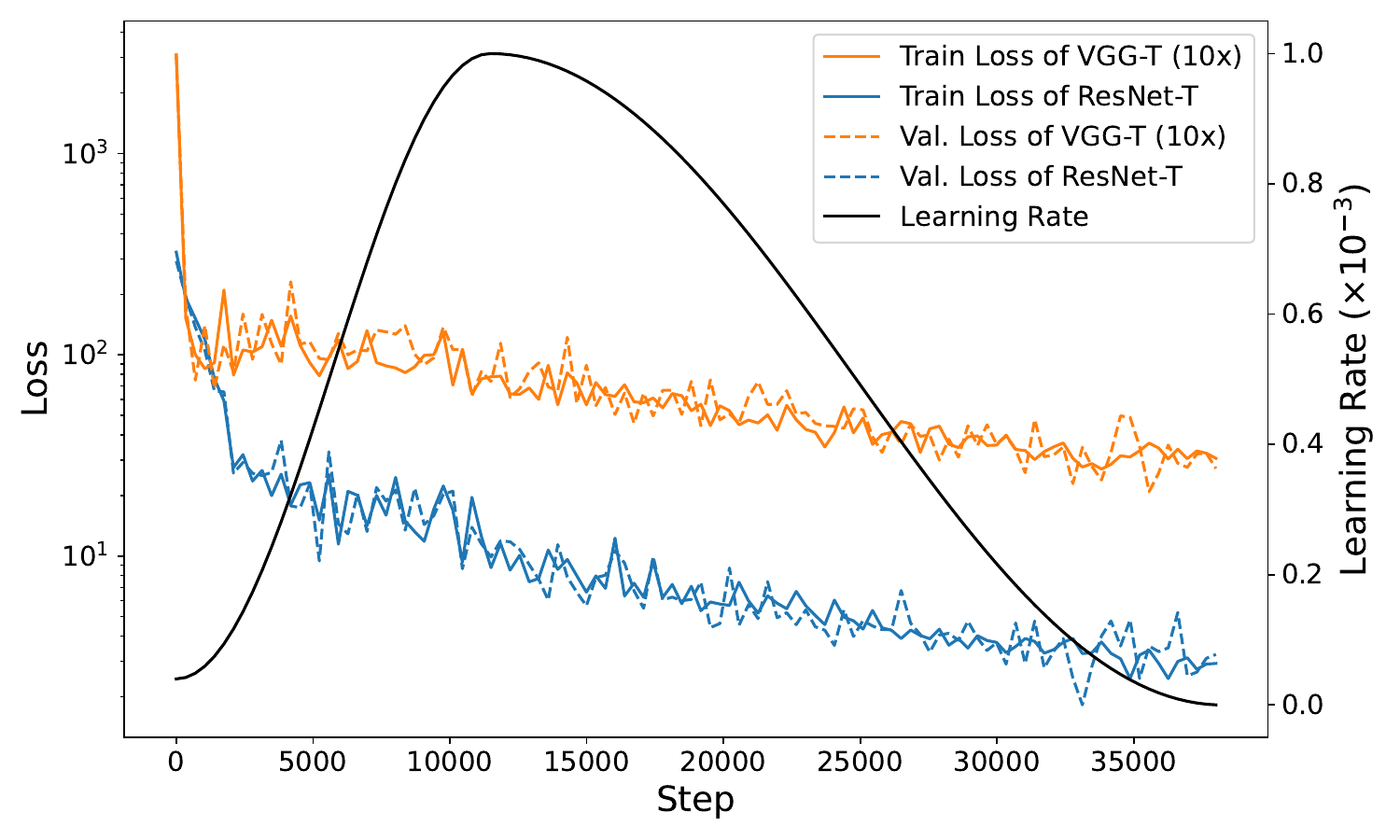}
    \caption{Training loss and learning rate curves for the VGG-T and ResNet-T models. To improve visualization, the training and validation loss values of VGG-T are multiplied by 10 (labeled as 10×) so that their trends can be more easily distinguished from those of ResNet-T, given their originally similar magnitudes.}
    \label{fig:loss_lr}
\end{figure}

Throughout the training process, ResNet-T utilizes the same training settings as those applied to VGG-T, ensuring consistency while allowing us to explore the benefits of depth via residual connections. This approach facilitates better performance in reconstruction tasks and fosters a more stable training process, ultimately contributing to improved accuracy and efficiency in the TAO experiment. Figure~\ref{fig:loss_lr} depicts the training dynamics of VGG-T and ResNet-T, showing the progression of the loss function and the learning rate schedule over the course of training, suggesting that both models converge smoothly without any evident signs of overfitting.

\section{Results and discussions}
\label{sec: R&D}
\begin{figure}[htbp]
    \centering
    \begin{subfigure}[b]{0.32\textwidth}
        \centering
        \includegraphics[width=\textwidth]{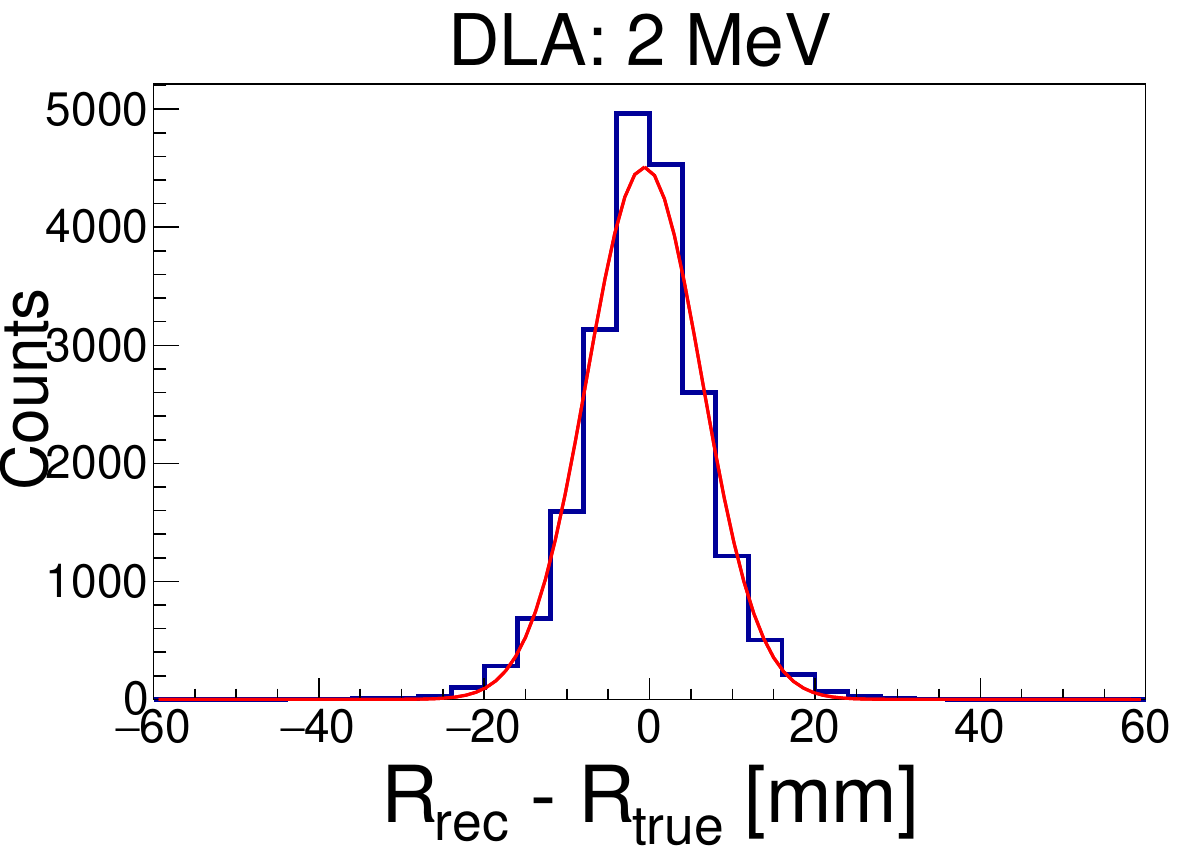}
        \caption{}
        \label{fig:DLA_2MeV}
    \end{subfigure}
    \hfill
    \begin{subfigure}[b]{0.32\textwidth}
        \centering
        \includegraphics[width=\textwidth]{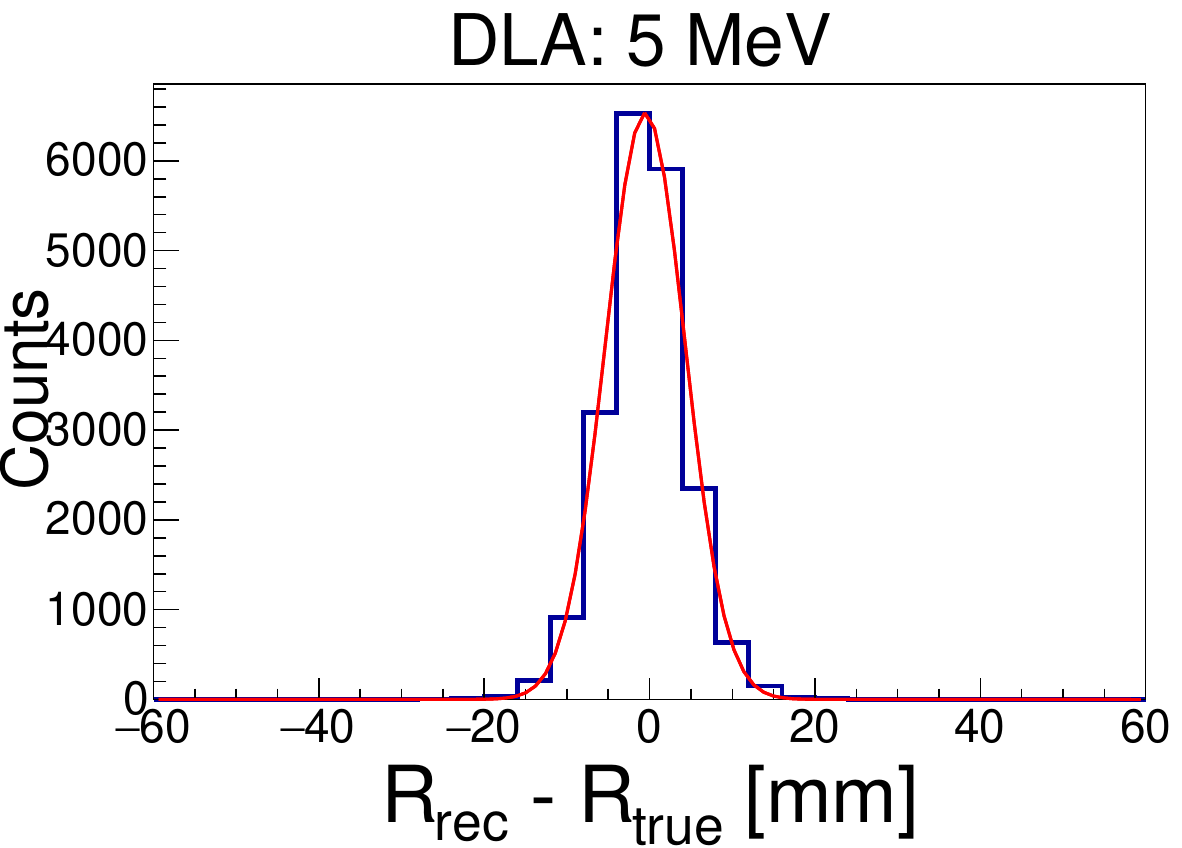}
        \caption{}
        \label{fig:DLA_5MeV}
    \end{subfigure}
    \hfill
    \begin{subfigure}[b]{0.32\textwidth}
        \centering
        \includegraphics[width=\textwidth]{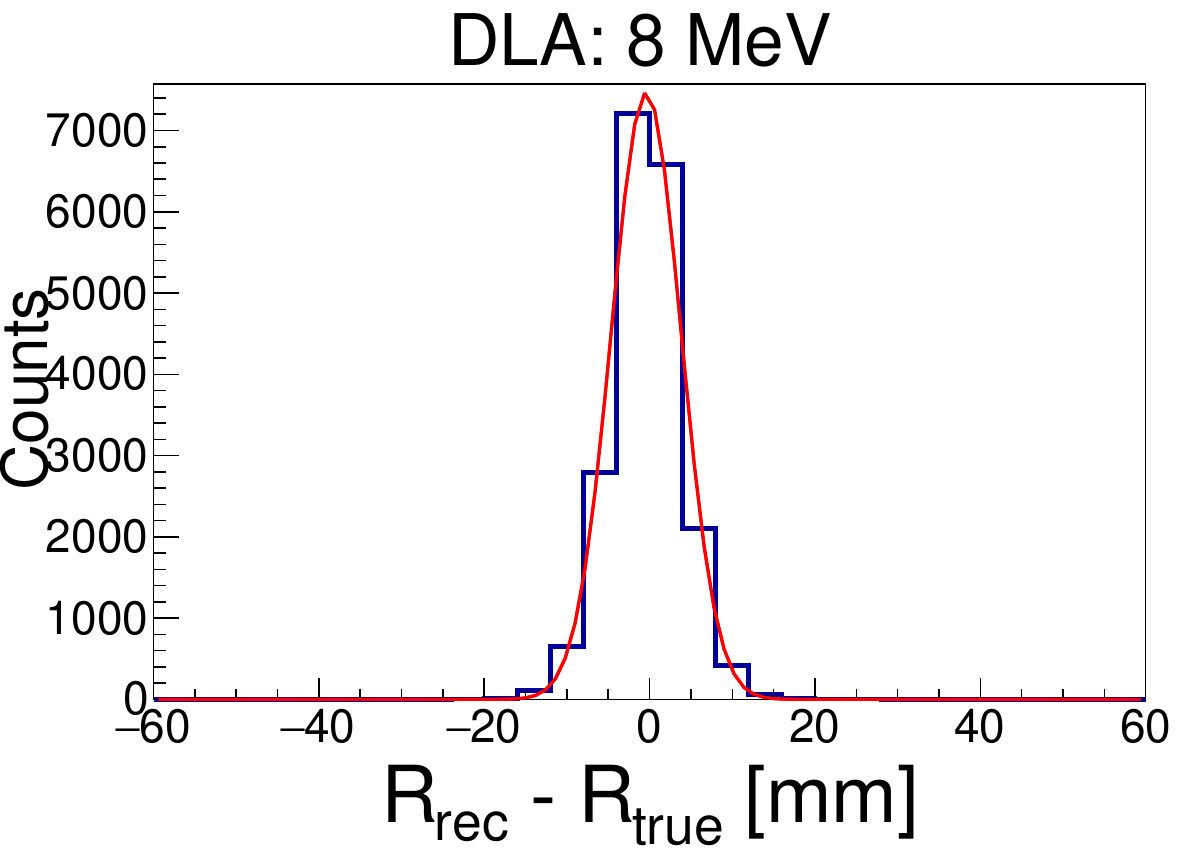}
        \caption{}
        \label{fig:DLA_8MeV}
    \end{subfigure}
    \begin{subfigure}[b]{0.32\textwidth}
        \centering
        \includegraphics[width=\textwidth]{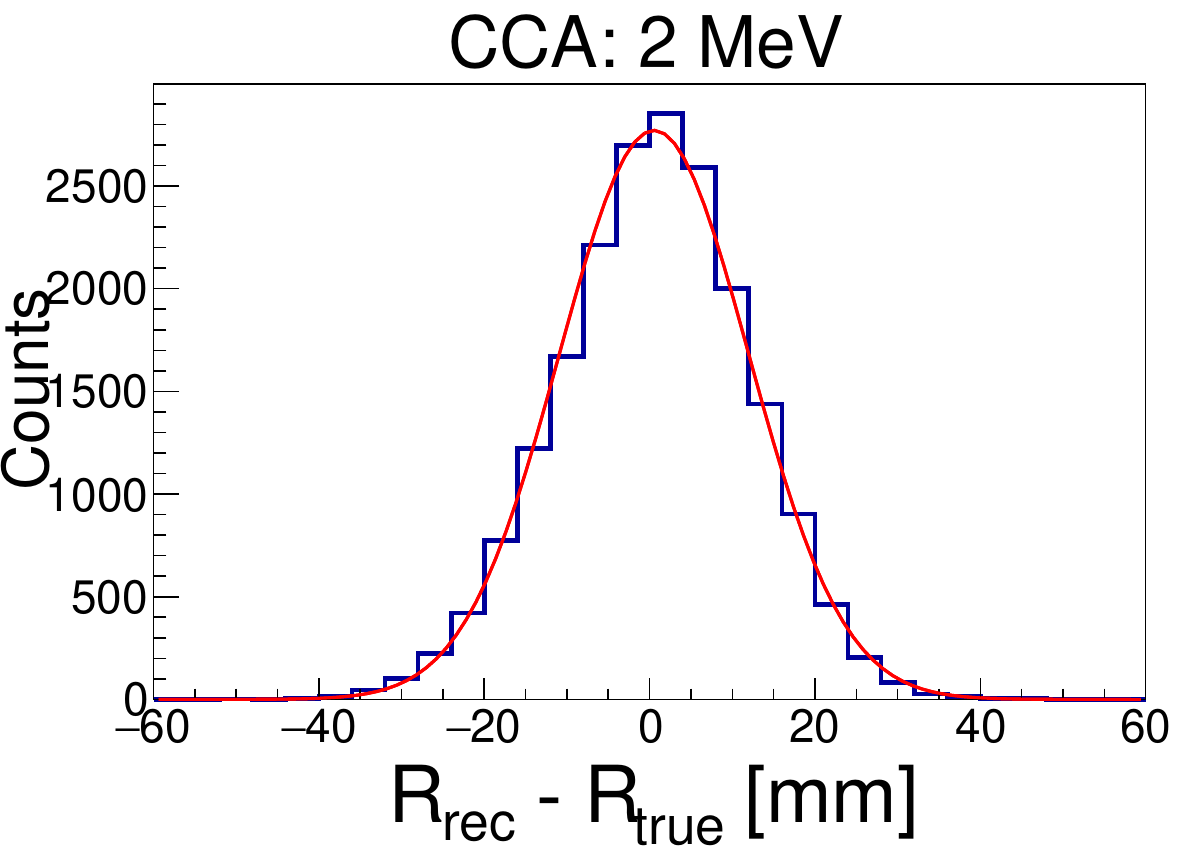}
        \caption{}
        \label{fig:CCA_2MeV}
    \end{subfigure}
    \hfill
    \begin{subfigure}[b]{0.32\textwidth}
        \centering
        \includegraphics[width=\textwidth]{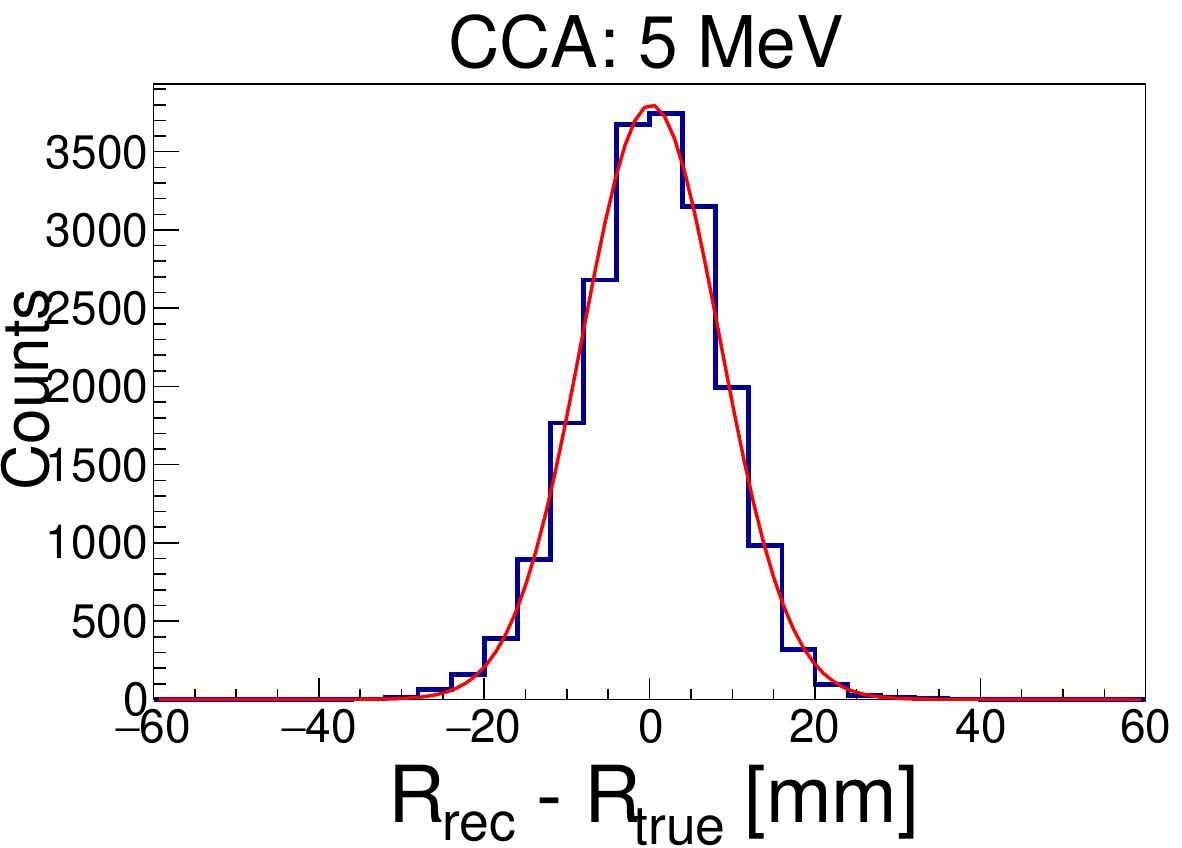}
        \caption{}
        \label{fig:CCA_5MeV}
    \end{subfigure}
    \hfill
    \begin{subfigure}[b]{0.32\textwidth}
        \centering
        \includegraphics[width=\textwidth]{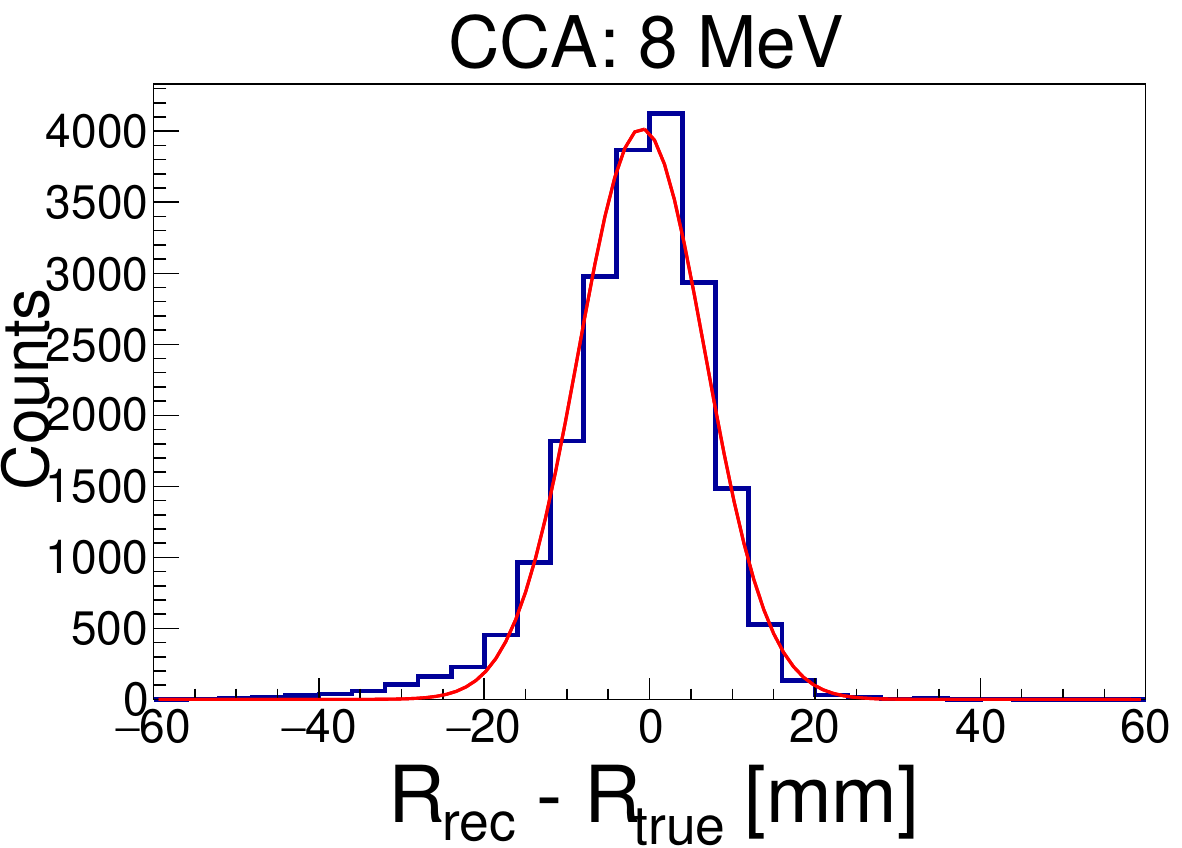}
        \caption{}
        \label{fig:CCA_8MeV}
    \end{subfigure}
\caption{Residual distributions within the full volume ($\rm R < 900~mm$) for DLA and CCA reconstructions (in electrons). The blue histograms and red curves show the true residual distributions and their corresponding Gaussian fits, respectively. Positrons and gammas exhibit similar behavior.}
\label{fig:energy_histogram}
\end{figure}

This section discusses the reconstruction performance of different algorithms. Leveraging its powerful nonlinear fitting capability, DLA produces reconstruction residuals that exhibit no obvious skewness or long tails, as shown in Fig.~\ref{fig:energy_histogram} (a)–(c). However, for high-energy particles, the residuals from CCA reconstruction across the full volume exhibit slight tails, deviating from a pure Gaussian distribution, as shown in Fig.~\ref{fig:energy_histogram} (e)–(f). This phenomenon is primarily caused by increased energy leakage near the spherical shell boundary for high-energy particles, which in turn leads to more non-Gaussian deviations in the residuals. Currently, CCA cannot fully mitigate this effect. To ensure a fair and consistent comparison, we employ the mean and standard deviation of the residuals as the evaluation metrics for reconstruction bias and resolution in this work, respectively.

\subsection{Performance improvement of optimized CCA}
\label{subsec:impacts4CCA}
In Section \ref{sec:improvements}, we introduced optimizations to CCA targeting dual-opening, $R$ reconstruction nonlinearity, and electronic effects in the TAO CD to enhance reconstruction performance. Since the improvements to the reconstruction of the angles $\theta$ and $\phi$ are negligible, this section is dedicated solely to systematically evaluating and discussing the impact of these optimizations on the $R$ reconstruction.

\begin{figure}[h]
    \centering
    \begin{subfigure}[b]{0.49\textwidth}
        \includegraphics[width=1\textwidth]{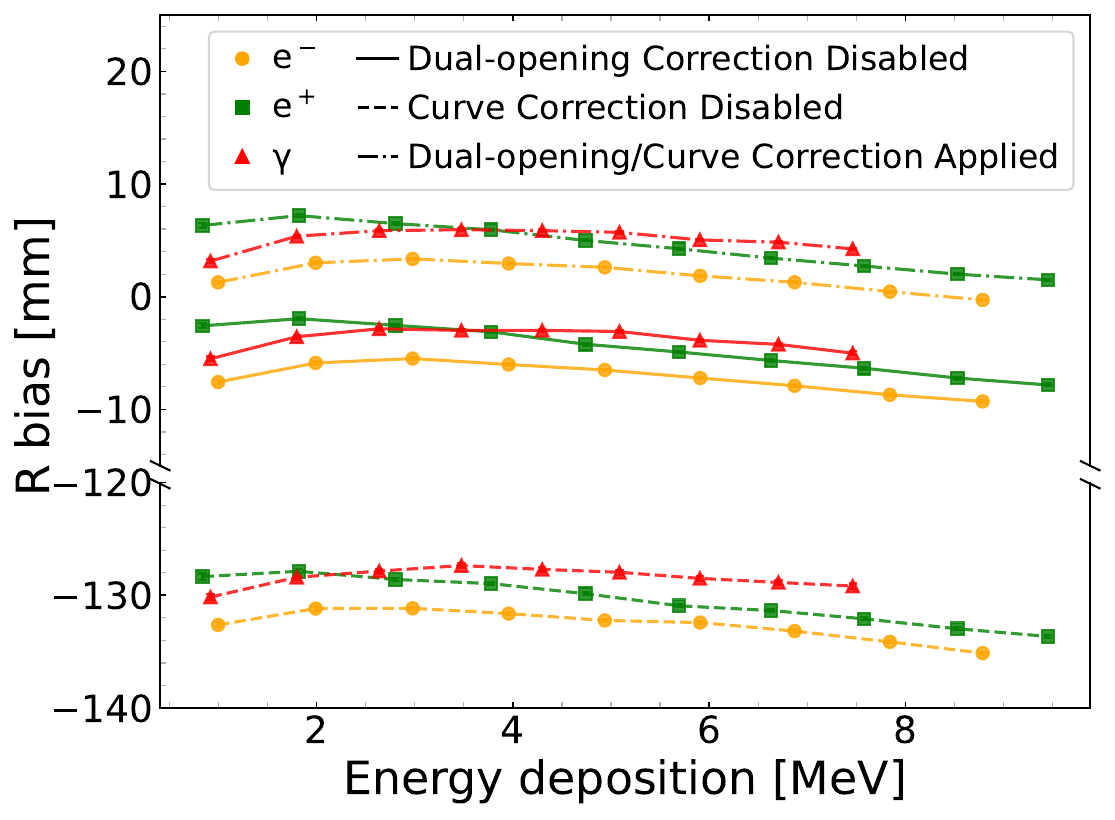}
    \caption{}
    \label{fig: correctionBias}
    \end{subfigure}
    \begin{subfigure}[b]{0.49\textwidth}
        \includegraphics[width=1\textwidth]{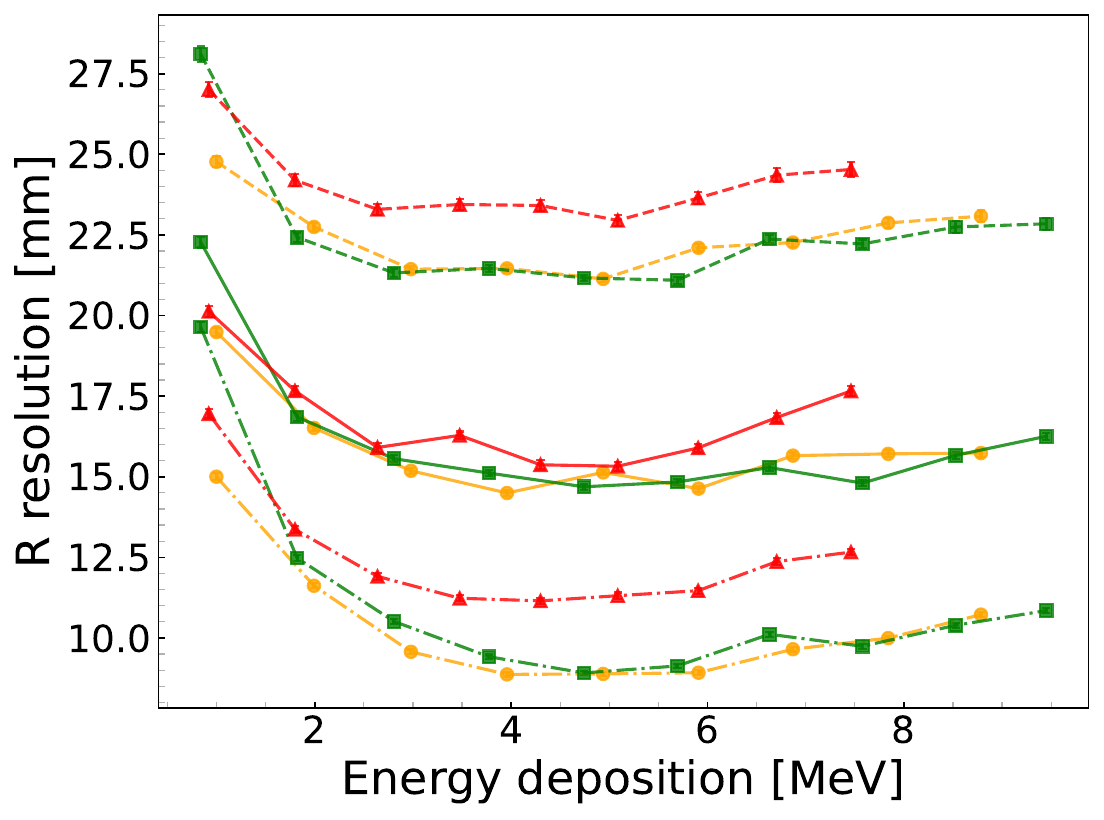}
        \caption{}
        \label{fig: correctionResolution}
    \end{subfigure}
    \caption{The $R$ reconstruction performance of the CCA under various corrections as a function of energy deposition. The applied corrections include dual-opening and curve corrections, with the latter implemented using the SCC fitting scheme. (a) and (b) show bias and resolution distributions at different energies, respectively. }
    \label{fig: correction}
\end{figure}

The impact of various corrections on the CCA reconstruction results is illustrated in Fig.~\ref{fig: correction}. Notably, the curve correction for the $R$ reconstruction nonlinearity has a particularly significant influence; after correction, the bias range for electrons, positrons, and gamma particles is optimized from [$-135.13\pm 0.14$ mm, $-127.36 \pm 0.20$ mm] to [$-0.28\pm 0.09$ mm, $7.2\pm 0.12$ mm]. The resolution range improves from [$21.13\pm 0.10$ mm, $28.5\pm 0.12$ mm] to [$8.87\pm 0.06$ mm, $19.6\pm 0.15$ mm]. In contrast, the dual-opening correction shows limited bias improvement yet substantially reduces photon leakage for all three particle types. Following this correction, the bias shifts from negative to positive, and the resolution is refined from [$14.49\pm 0.09$ mm, $22.27\pm 0.16$ mm] to [$8.87\pm 0.06$ mm, $19.6\pm 0.15$ mm].

\begin{figure}[h]
    \centering
    \begin{subfigure}[b]{0.49\textwidth}
        \includegraphics[width=1\textwidth]{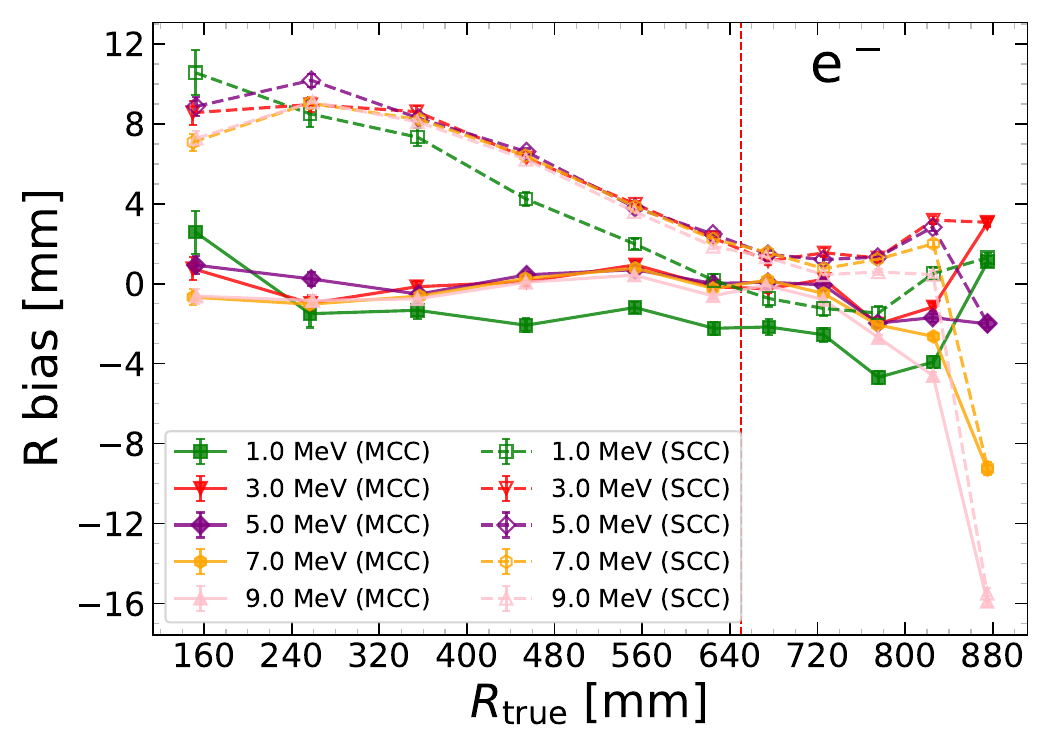}
    \caption{}
    \label{fig: e-CurveCorrectionBias}
    \end{subfigure}
    \begin{subfigure}[b]{0.49\textwidth}
        \includegraphics[width=1\textwidth]{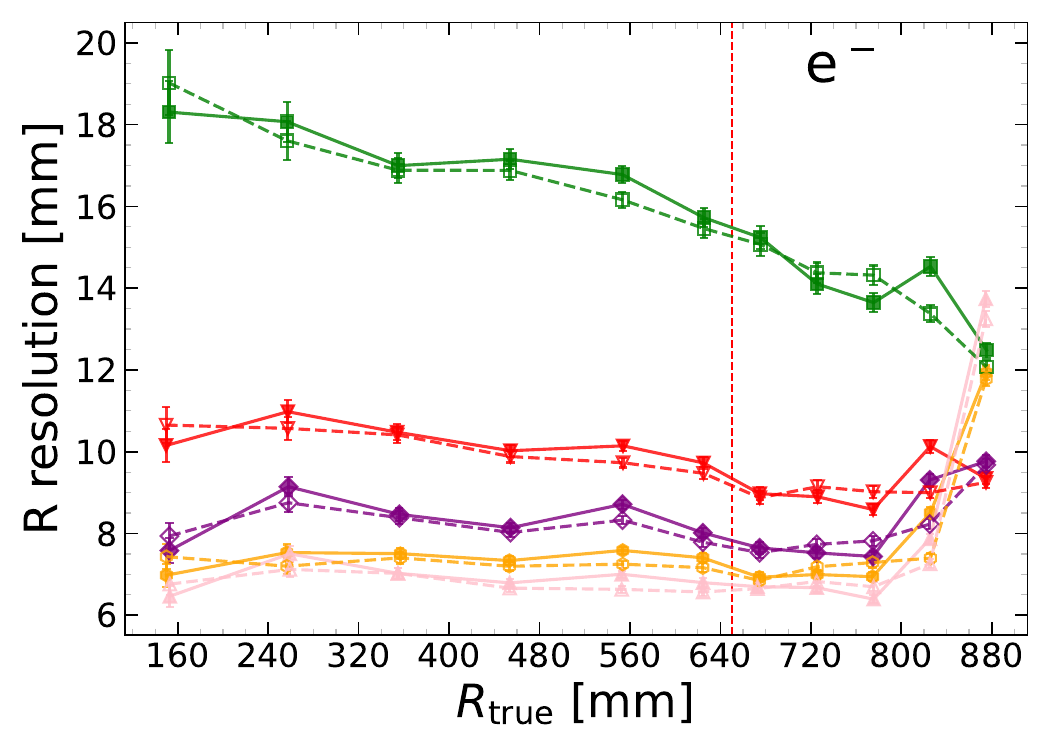}
        \caption{}
        \label{fig: e-CurveCorrectionResolution}
    \end{subfigure}
    \begin{subfigure}[b]{0.49\textwidth}
        \includegraphics[width=1\textwidth]{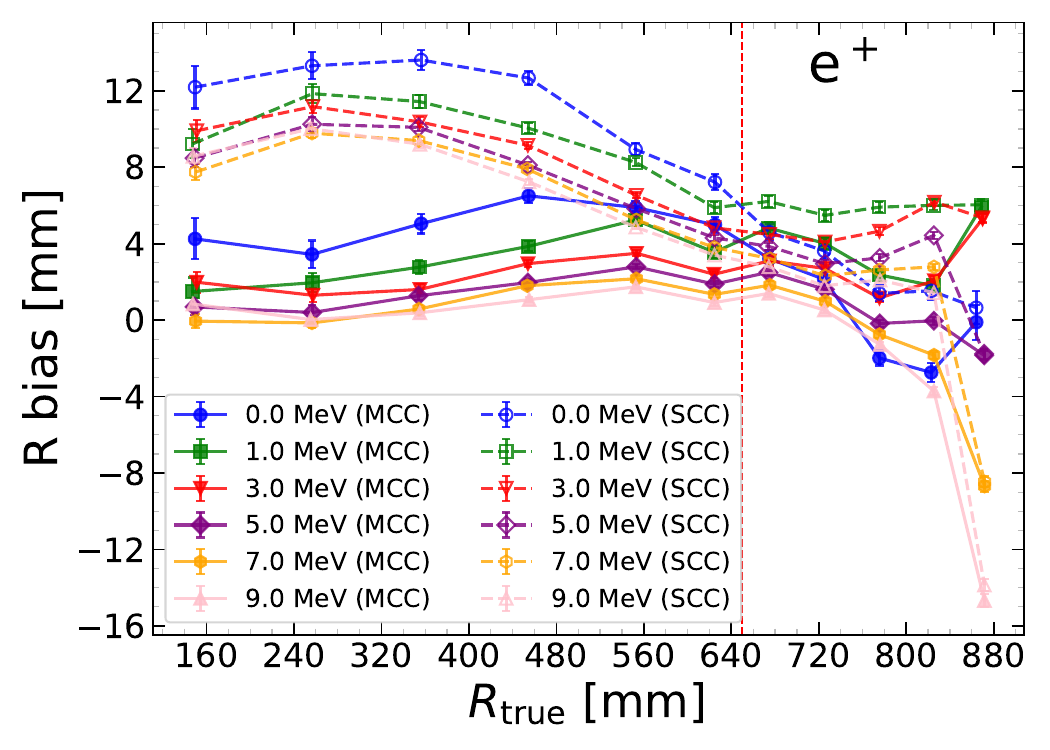}
    \caption{}
    \label{fig: e+CurveCorrectionBias}
    \end{subfigure}
    \begin{subfigure}[b]{0.49\textwidth}
        \includegraphics[width=1\textwidth]{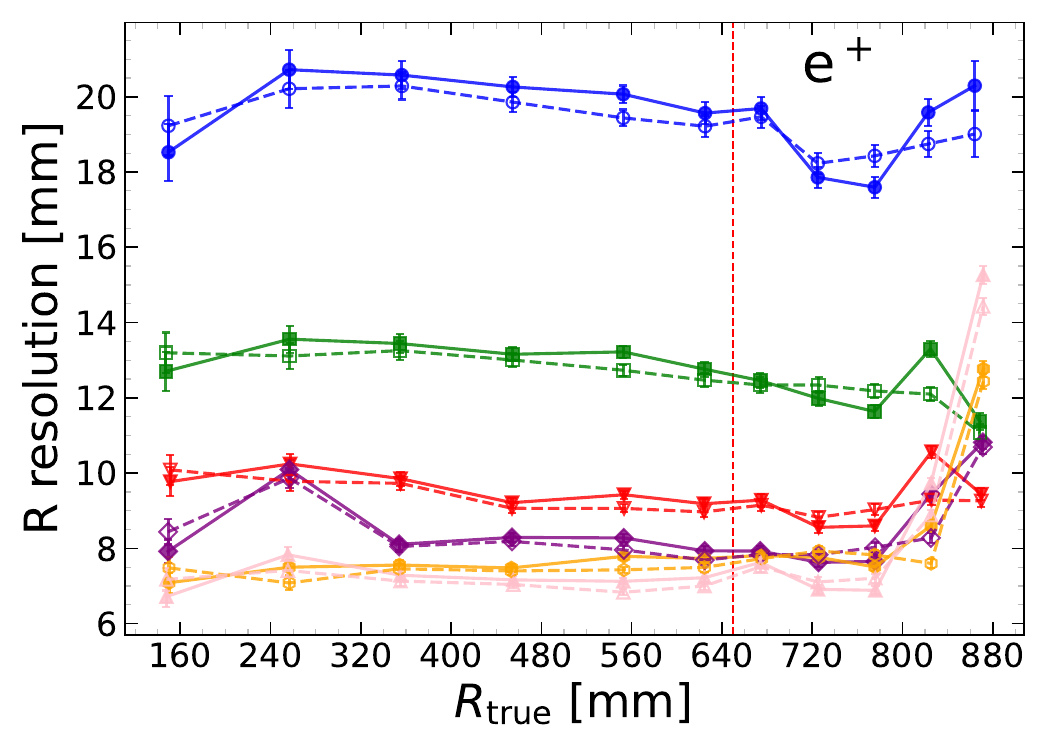}
        \caption{}
        \label{fig: e+CurveCorrectionResolution}
    \end{subfigure}
    \begin{subfigure}[b]{0.49\textwidth}
        \includegraphics[width=1\textwidth]{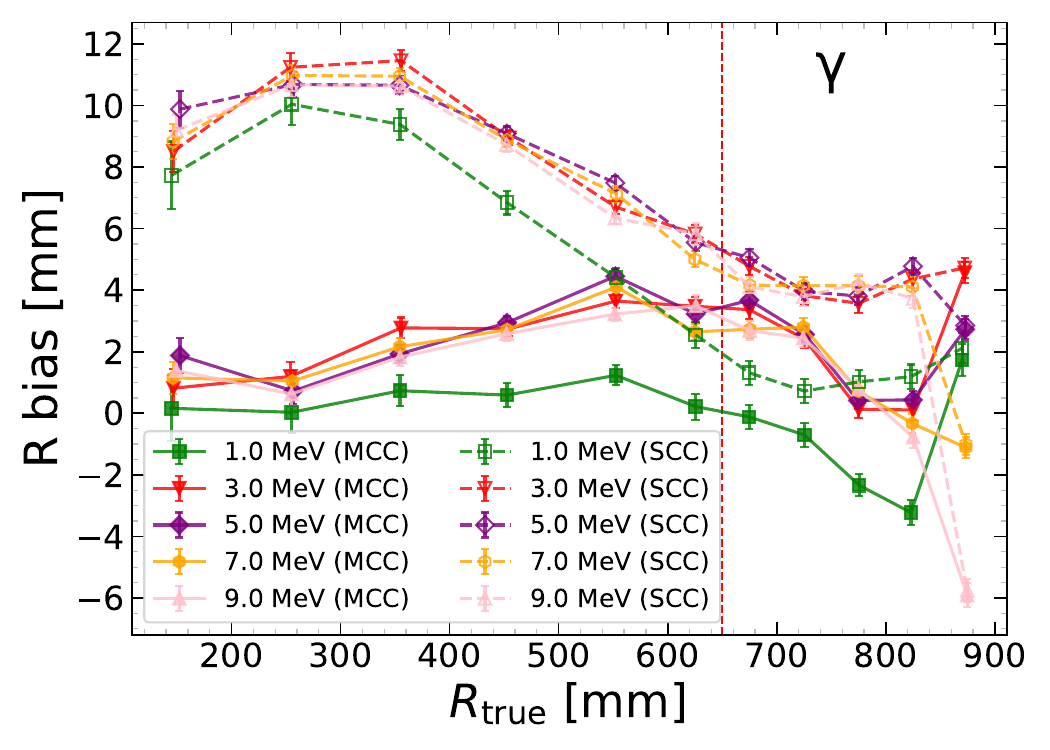}
    \caption{}
    \label{fig: GammaCurveCorrectionBias}
    \end{subfigure}
    \begin{subfigure}[b]{0.49\textwidth}
        \includegraphics[width=1\textwidth]{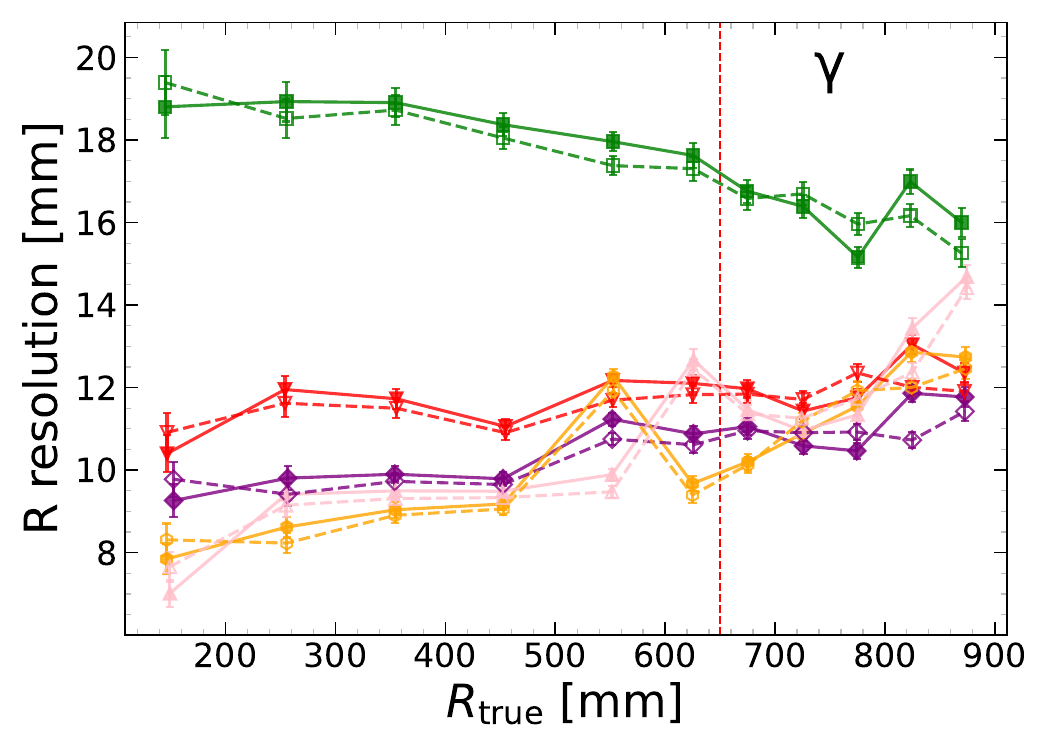}
        \caption{}
        \label{fig: GammaCurveCorrectionResolution}
    \end{subfigure}
    \caption{The $R$ reconstruction performance of the CCA regarding energy deposition position. The red dashed line indicates the boundary of the fiducial volume, with $\mathit{{R}_\text{true}} = 650~\text{mm}$. The left column shows the reconstructed $R$ bias, while the right column shows the reconstructed $R$ resolution. All results for the dual-opening have been corrected.}
    \label{fig: CurveCorrection}
\end{figure}

The SCC and MCC fitting schemes are compared in Fig.~\ref{fig: CurveCorrection}. Across the entire detector region, both schemes exhibit minimal resolution differences for different particles. However, within the fiducial volume region, the bias of the SCC fitting scheme increases significantly as the radius decreases, surpassing that of the MCC fitting scheme. Outside this region, the bias difference between the two schemes is minor, but near the boundary, the SCC scheme's results deteriorate again. This observation is consistent with the analysis in Section~\ref{sec:improvements}, where a lower density of calibration points at smaller radius regions leads to reduced correction accuracy for the SCC fitting scheme.

\begin{figure}[h]
    \centering
    \begin{subfigure}[b]{0.49\textwidth}
        \includegraphics[width=1\textwidth]{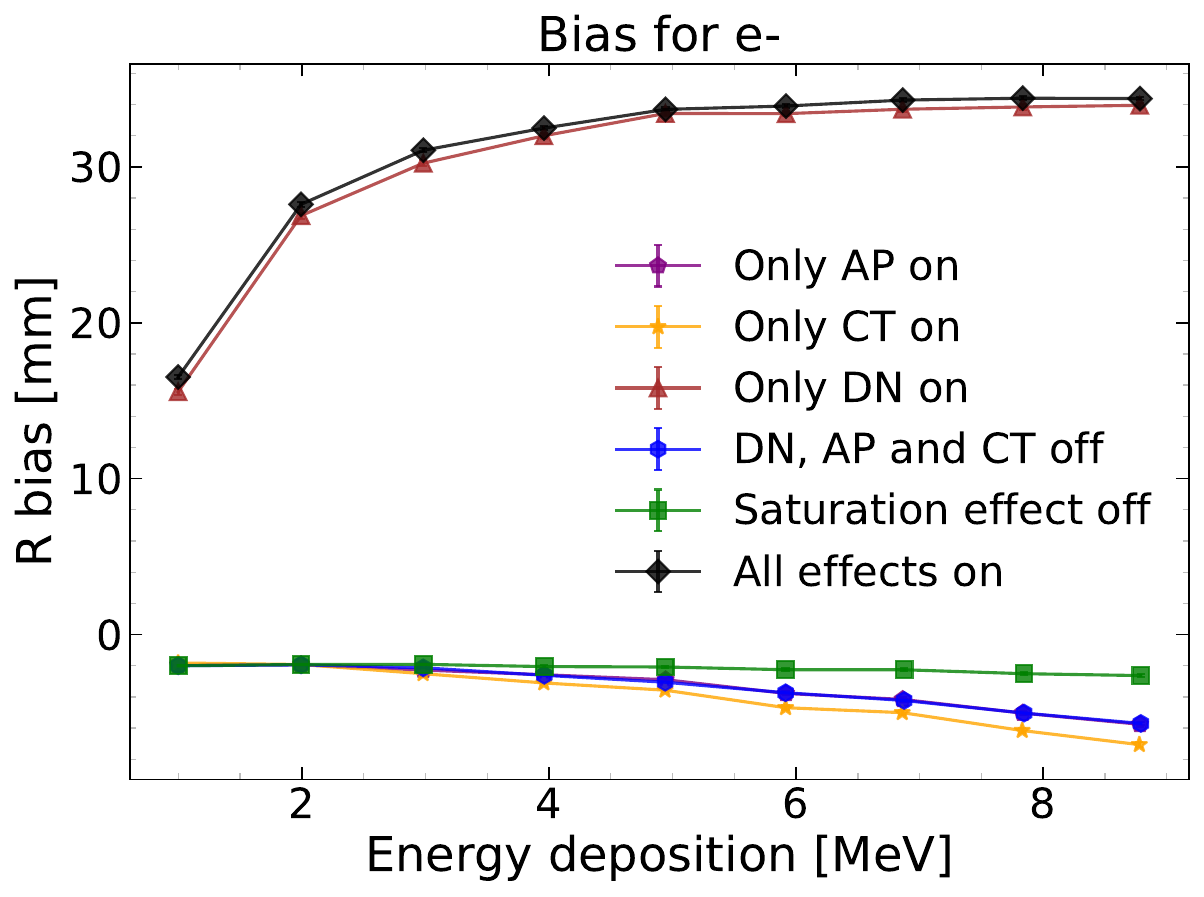}
    \caption{}
    \label{fig: SiPMEffBias}
    \end{subfigure}
    \begin{subfigure}[b]{0.49\textwidth}
        \includegraphics[width=1\textwidth]{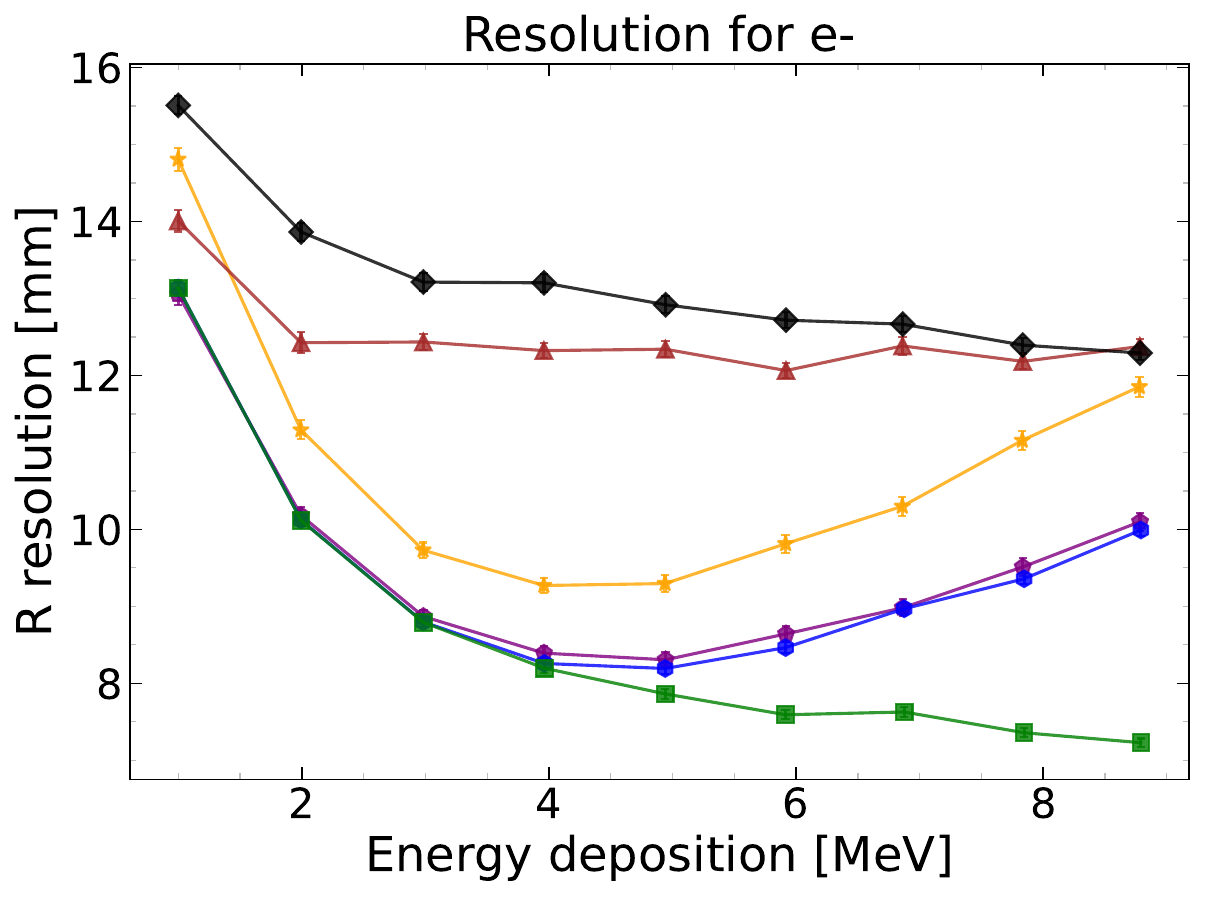}
        \caption{}
        \label{fig: SiPMEffResolution}
    \end{subfigure}
    \caption{The impact of electronic effects on the $R$ reconstruction results of electron events. These effects have a similar influence on positrons and gamma particles. (a) and (b) show bias and resolution distributions at different energies, respectively.}
    \label{fig: SiPMeff}
\end{figure}

The contribution of various electronic effects to the reconstruction results using the CCA is presented in Fig.~\ref{fig: SiPMeff}. The introduction of DN significantly degrades $R$ reconstruction accuracy because the charge induced by this background distorts the spatial charge distribution, leading to a systematic bias in vertex reconstruction, especially in low-energy events. Compared to the results without electronic effects, the addition of CT alone causes both the bias and resolution to deteriorate, and the difference in bias is more pronounced for higher-energy particles. This phenomenon arises because higher-energy particles produce more hits on the SiPMs, increasing the CT probability. Additionally, the saturation effect of SiPM readout channels is another critical factor affecting $R$ reconstruction accuracy. Saturation emerges when the readout channels reach their counting limit, causing excess counts to be discarded. For higher-energy particles, which yield more hits on SiPMs, the truncation and loss of signals due to saturation are more significant, thus posing a greater challenge for the vertex reconstruction of such particles. The AP exerts the most negligible effect on the $R$ reconstruction accuracy among all effects, given that its charge count is far less than that of the signal.

\subsection{Comparison of reconstruction performance between CCA and DLA}
\label{subsec:comparion2algorithms}
\begin{figure}
    \centering
    \includegraphics[width=0.6\linewidth]{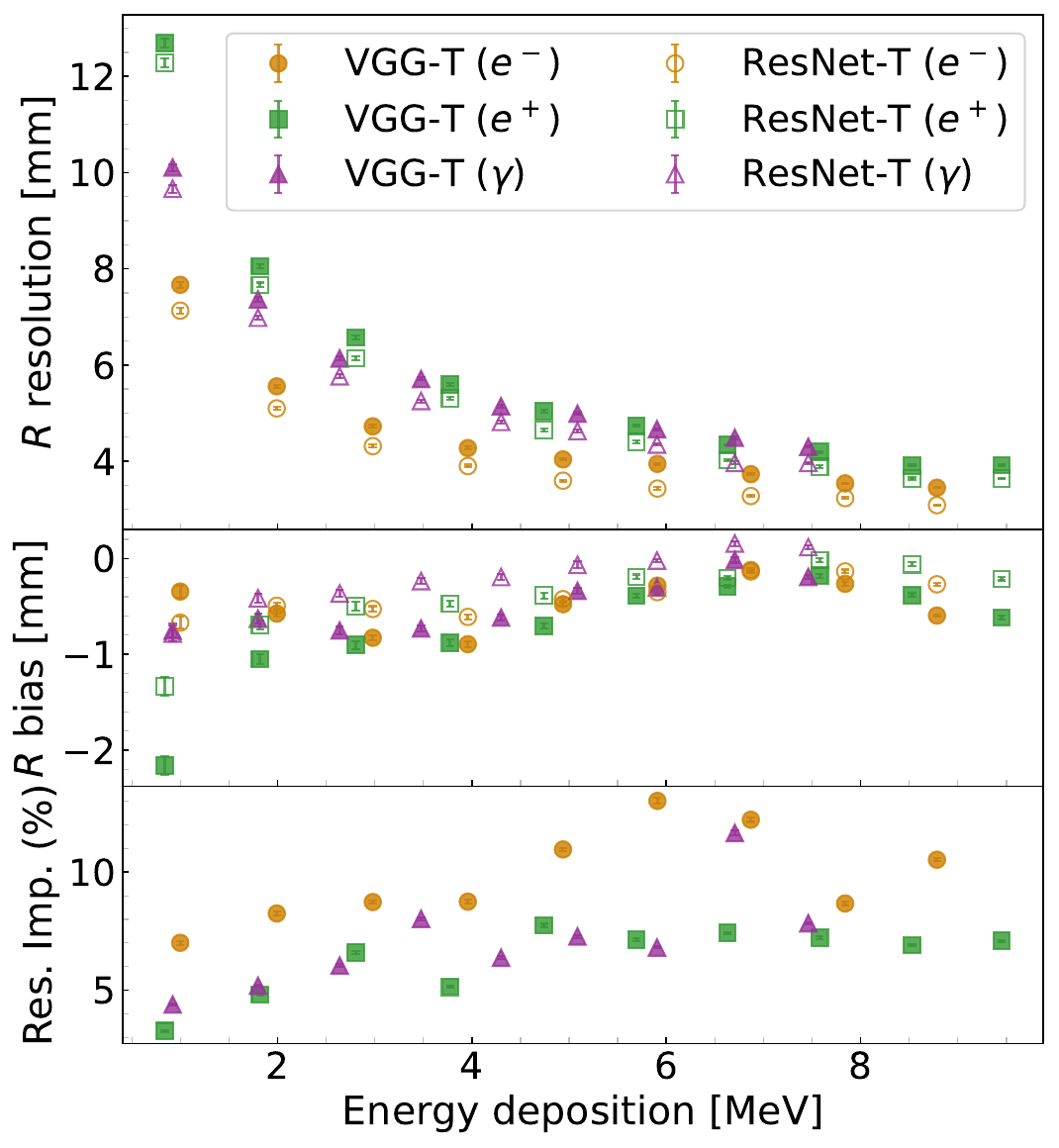}
    \caption{The performance of DLA in $R$ reconstruction. The upper panel presents the  $R$ reconstruction resolution of VGG-T and ResNet-T, while the middle panel shows their reconstruction bias. The lower panel demonstrates the improvement in $R$ resolution achieved by ResNet-T compared to VGG-T.}
    \label{fig:DL-R}
\end{figure}

The reconstruction results of DLA in $R$ reconstruction are shown in Fig.~\ref{fig:DL-R}. In terms of bias, the outcomes for all three particle types are generally controlled within 1 mm. However, positrons with initial kinetic energies below 2 MeV exhibit a larger bias exceeding 1 mm. This phenomenon arises because the gamma rays produced during the annihilation of low-energy positrons carry a higher proportion of energy, leading to a more dispersed distribution of energy deposition positions and consequently increasing the reconstruction error. Regarding reconstruction resolution, the $R$ resolution values demonstrate a decreasing trend as the deposited energy increases. Among the particles, electrons exhibit the best resolution performance, benefiting from their more concentrated energy deposition under the same deposited energy conditions, which enables more precise vertex localization and reduces statistical errors. Comparing the results of the two network architectures, ResNet-T, with its deeper network layers, shows improvements in $R$ reconstruction bias and resolution, particularly notable resolution enhancement. It is worth noting that the improvement in electron resolution is the most significant, reaching up to 13\%, highlighting the substantial optimization effect of this network structure on electron reconstruction performance. In addition, we conducted both generalization and robustness tests on both models, and the test results indicate that they both meet the reconstruction accuracy requirements.

\begin{figure}
    \centering
    \includegraphics[width=0.6\linewidth]{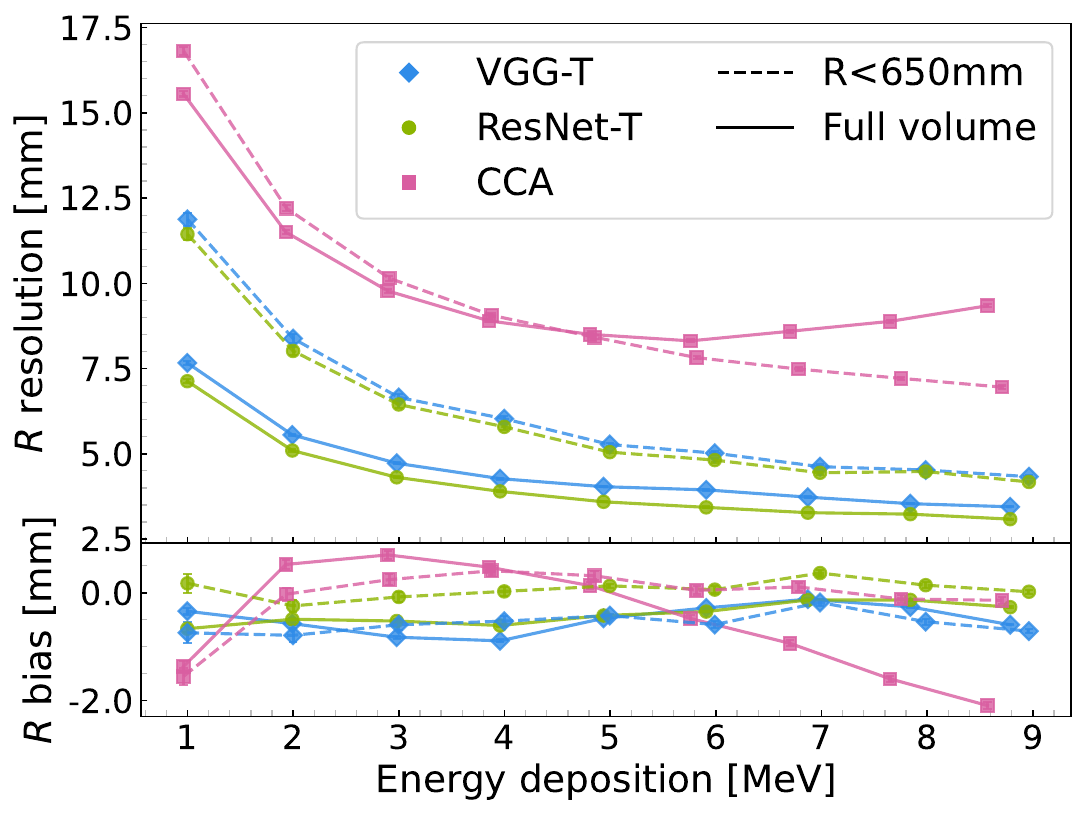}
    \caption{Comparison of $R$ reconstruction performance between DLA and CCA for electrons. All data are based on the full electronic effects simulation.}
    \label{fig:DL-CCA-R}
\end{figure}
\begin{figure}[h]
    \centering
    \begin{subfigure}[b]{0.49\textwidth}
        \includegraphics[width=1\textwidth]{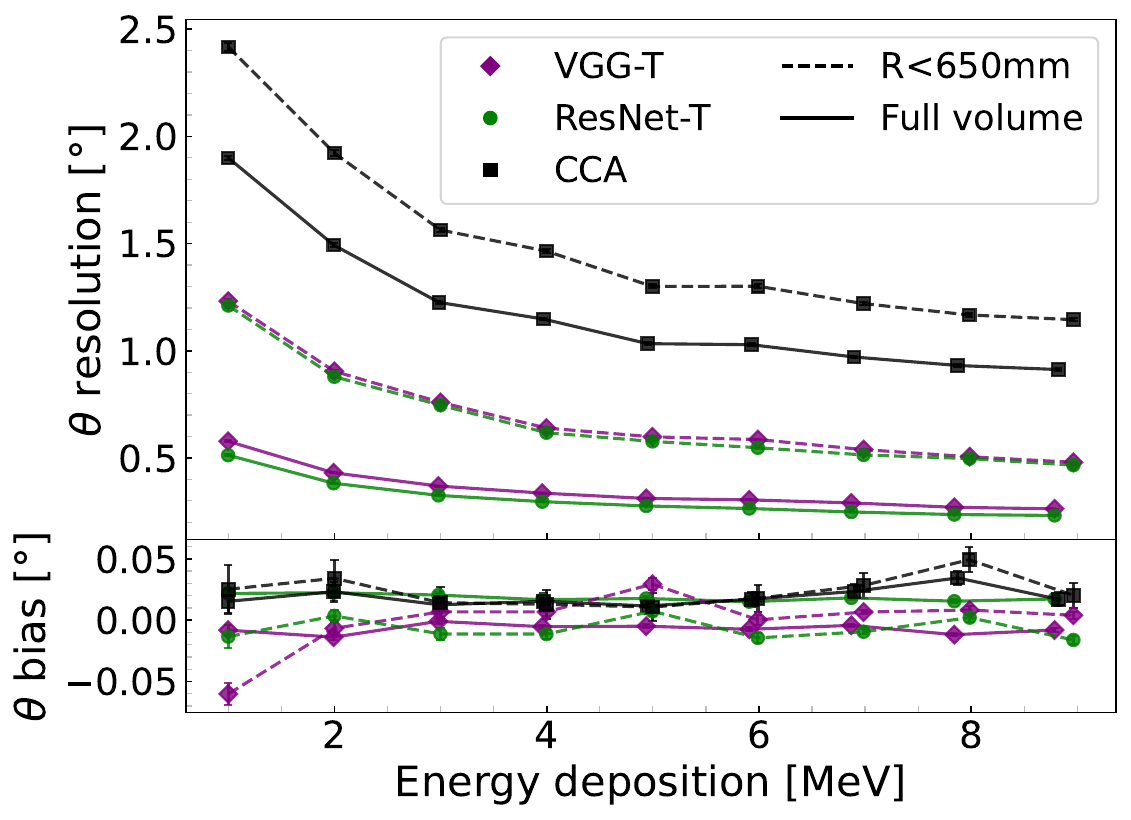}
    \caption{}
    \label{fig: Theta}
    \end{subfigure}
    \begin{subfigure}[b]{0.482\textwidth}
        \includegraphics[width=1\textwidth]{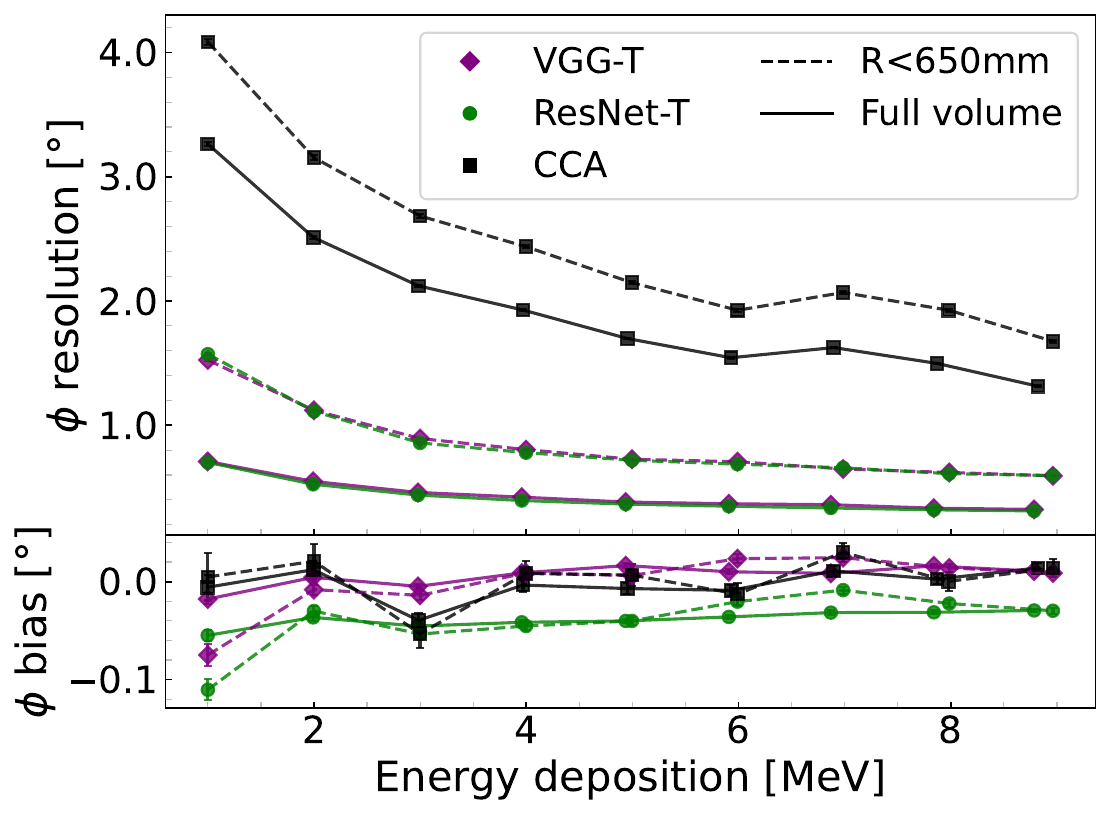}
        \caption{}
        \label{fig: Phi}
    \end{subfigure}
    \caption{Electron angular reconstruction performance. Results for positrons and gamma rays show only minor numerical differences. (a) describes the bias and resolution distribution of $\theta$ reconstruction. (b) describes the bias and resolution distribution of $\phi$ reconstruction.}
    \label{fig:Angle}
\end{figure}

Figure~\ref{fig:DL-CCA-R} compares electron reconstruction performance between the CCA and DLA, revealing several key advantages of the latter. The DLA yields smaller biases with more stable trends than the CCA and delivers markedly better resolution under all conditions. Although the DLA's resolution slightly degrades within the fiducial volume ($R < 650\,$mm), this reflects its superior handling of boundary events. These edge events, despite involving more complex electronic effects, benefit from a more concentrated charge distribution on the SiPM channels (as shown in Fig.~\ref{fig: Charge2D}) and enhanced distinctiveness of image features, enabling the DLA to handle boundary events more effectively. This variation in performance suggests that DLA is particularly effective at processing spatially localized signal patterns that are characteristic of the detector edge regions. When using the DLA, comparable improvements are observed for gamma-ray and positron events.

The reconstruction performance in the $\mathrm{\theta}$ and $\mathrm{\phi}$ directions for the DLA and CCA is compared in Fig. \ref{fig:Angle}. Both algorithms exhibit similar reconstruction biases, but the DLA provides significantly better resolution than the CCA. Notably,  employing a deeper ResNet-T network structure in the DLA, instead of VGG-T, does not substantially enhance the reconstruction performance in the $\mathrm{\theta}$ and $\mathrm{\phi}$ directions, indicating that angular reconstruction is not sensitive to network depth. Additionally, both algorithms demonstrate poorer reconstruction performance for events occurring within the fiducial volume, yet perform better for edge events. This is attributed to the fact that edge events feature more concentrated SiPM hits and reduced angular dispersion, contributing to enhanced reconstruction accuracy.

\section{Summary}
\label{sec: summary}
This work proposes two distinct vertex reconstruction approaches, CCA and DLA, for the TAO experiment. Through extensive numerical simulations, we systematically evaluate and compare the performance of both methods in vertex reconstruction for electrons, gamma rays, and positrons. 

In the case of the CCA, we introduced innovative corrections, including dual-opening and curve correction techniques, by carefully analyzing the TAO detector's intrinsic characteristics and electronic effects. The CCA achieves excellent reconstruction performance: at 1 MeV energy deposition, the $R$ resolution is better than $20~\mathrm{mm}$ (bias $<5~\mathrm{mm}$), $\mathrm{\theta}$ resolution better than $2.5~\degree$ (bias $<0.1~\degree$), and $\mathrm{\phi}$ resolution better than $4~\degree$ (bias $<0.1~\degree$) across the full detector volume for all three particle types. 
Two deep network structures, VGG-T and ResNet-T, were constructed for the DLA. Experimental results indicate that the ResNet-T network demonstrates superior reconstruction accuracy at 1 MeV energy deposition, achieving $R$ resolution $<12~\mathrm{mm}$ (bias $<1.3~\mathrm{mm}$), $\mathrm{\theta}$ resolution $<1.6~\degree$ (bias $<0.05~\degree$), and $\mathrm{\phi}$ resolution $<1.5~\degree$ (bias $<0.05~\degree$). While improvements in $\mathrm{\theta}$ and $\mathrm{\phi}$ reconstruction with ResNet-T compared to VGG-T are limited, there is a significant enhancement in $R$ resolution, with improvements of 9.8\%, 6.5\%, and 7.1\% for electrons, positrons, and gamma events, respectively.

A comprehensive comparison indicates that the overall performance of the DLA exceeds that of the CCA. However, during the initial stages of experimentation, the CCA proves to be more practical when real data is limited. For detectors like TAO, which have a small volume ($R \sim 0.9~\rm m$) leading to low time sensitivity, but also feature high photon sensor coverage, the corrected CCA can still achieve high reconstruction accuracy without relying on timing information. Moreover, the CCA is much less time-consuming than the traditional time-based and likelihood algorithms, as it does not require iteration or gradient minimization, and can also provide initial values to improve the efficiency and convergence of these algorithms.
Future research will focus on continuously optimizing the deep learning models trained on simulated data by leveraging calibration data obtained from experiments to further enhance the applicability of the DLA to real experimental data.

\acknowledgments
We thank all members of the TAO group for fruitful discussions. 
This work was supported by the Fundamental Research Funds for the Central Universities, Sun Yat-sen University (Grant No. 24qnpy125), the China Postdoctoral Science Foundation (Grant No. 2024M753715), and the Guangdong Basic and Applied Basic Research Foundation (Grant No. 2023B1515120030).

\newpage
\bibliographystyle{unsrt}
\bibliography{newrefs}

\begin{thebibliography}{10}

\bibitem{RENO:2015ksa}
J.~H. Choi et~al.
\newblock {Observation of Energy and Baseline Dependent Reactor Antineutrino
  Disappearance in the RENO Experiment}.
\newblock {\em Phys. Rev. Lett.}, 116(21):211801, 2016.

\bibitem{DayaBay:2015lja}
Feng~Peng An et~al.
\newblock {Measurement of the Reactor Antineutrino Flux and Spectrum at Daya
  Bay}.
\newblock {\em Phys. Rev. Lett.}, 116(6):061801, 2016.
\newblock [Erratum: Phys.Rev.Lett. 118, 099902 (2017)].

\bibitem{DoubleChooz:2015mfm}
Y.~Abe et~al.
\newblock {Measurement of \ensuremath{\theta}$_{13}$ in Double Chooz using
  neutron captures on hydrogen with novel background rejection techniques}.
\newblock {\em JHEP}, 01:163, 2016.

\bibitem{NEOS:2016wee}
Y.~J. Ko et~al.
\newblock {Sterile Neutrino Search at the NEOS Experiment}.
\newblock {\em Phys. Rev. Lett.}, 118(12):121802, 2017.

\bibitem{STEREO:2022nzk}
H.~Almaz\'an et~al.
\newblock {STEREO neutrino spectrum of $^{235}$U fission rejects sterile
  neutrino hypothesis}.
\newblock {\em Nature}, 613(7943):257--261, 2023.

\bibitem{Huber:2011wv}
Patrick Huber.
\newblock {On the determination of anti-neutrino spectra from nuclear
  reactors}.
\newblock {\em Phys. Rev. C}, 84:024617, 2011.
\newblock [Erratum: Phys.Rev.C 85, 029901 (2012)].

\bibitem{Mueller:2011nm}
Th.~A. Mueller et~al.
\newblock {Improved Predictions of Reactor Antineutrino Spectra}.
\newblock {\em Phys. Rev. C}, 83:054615, 2011.

\bibitem{Mention:2011rk}
G.~Mention, M.~Fechner, Th. Lasserre, Th.~A. Mueller, D.~Lhuillier, M.~Cribier,
  and A.~Letourneau.
\newblock {The Reactor Antineutrino Anomaly}.
\newblock {\em Phys. Rev. D}, 83:073006, 2011.

\bibitem{Seo:2014xei}
Seon-Hee Seo.
\newblock {New Results from RENO and The 5 MeV Excess}.
\newblock {\em AIP Conf. Proc.}, 1666(1):080002, 2015.

\bibitem{Huber:2016xis}
Patrick Huber.
\newblock {NEOS Data and the Origin of the 5 MeV Bump in the Reactor
  Antineutrino Spectrum}.
\newblock {\em Phys. Rev. Lett.}, 118(4):042502, 2017.

\bibitem{JUNO:2024jaw}
Angel Abusleme et~al.
\newblock {Potential to identify neutrino mass ordering with reactor
  antineutrinos at JUNO*}.
\newblock {\em Chin. Phys. C}, 49(3):033104, 2025.

\bibitem{RELICS:2024opj}
Chang Cai et~al.
\newblock {Reactor neutrino liquid xenon coherent elastic scattering
  experiment}.
\newblock {\em Phys. Rev. D}, 110(7):072011, 2024.

\bibitem{JUNO:2020ijm}
Angel Abusleme et~al.
\newblock {TAO Conceptual Design Report: A Precision Measurement of the Reactor
  Antineutrino Spectrum with Sub-percent Energy Resolution}.
\newblock 5 2020.

\bibitem{Xu:2022mdi}
Hangkun Xu et~al.
\newblock {Calibration strategy of the JUNO-TAO experiment}.
\newblock {\em Eur. Phys. J. C}, 82(12):1112, 2022.

\bibitem{DayaBay:2012aa}
F.~P. An et~al.
\newblock {A side-by-side comparison of Daya Bay antineutrino detectors}.
\newblock {\em Nucl. Instrum. Meth. A}, 685:78--97, 2012.

\bibitem{DayaBay:2016ggj}
Feng~Peng An et~al.
\newblock {Measurement of electron antineutrino oscillation based on 1230 days
  of operation of the Daya Bay experiment}.
\newblock {\em Phys. Rev. D}, 95(7):072006, 2017.

\bibitem{Li:2021oos}
Ziyuan Li et~al.
\newblock {Event vertex and time reconstruction in large-volume liquid
  scintillator detectors}.
\newblock {\em Nucl. Sci. Tech.}, 32(5):49, 2021.

\bibitem{KamLAND:2002uet}
K.~Eguchi et~al.
\newblock {First results from KamLAND: Evidence for reactor anti-neutrino
  disappearance}.
\newblock {\em Phys. Rev. Lett.}, 90:021802, 2003.

\bibitem{Borexino:2008gab}
G.~Alimonti et~al.
\newblock {The Borexino detector at the Laboratori Nazionali del Gran Sasso}.
\newblock {\em Nucl. Instrum. Meth. A}, 600:568--593, 2009.

\bibitem{DoubleChooz:2012gmf}
Y.~Abe et~al.
\newblock {Reactor electron antineutrino disappearance in the Double Chooz
  experiment}.
\newblock {\em Phys. Rev. D}, 86:052008, 2012.

\bibitem{Liu:2018fpq}
Qin Liu, Miao He, Xuefeng Ding, Weidong Li, and Haiping Peng.
\newblock {A vertex reconstruction algorithm in the central detector of JUNO}.
\newblock {\em JINST}, 13(09):T09005, 2018.

\bibitem{Huang:2022zum}
Gui-hong Huang, Wei Jiang, Liang-jian Wen, Yi-fang Wang, and Wu-Ming Luo.
\newblock {Data-driven simultaneous vertex and energy reconstruction for large
  liquid scintillator detectors}.
\newblock {\em Nucl. Sci. Tech.}, 34(6):83, 2023.

\bibitem{Liu:2024cxo}
Xuewei Liu, Wei Dou, Benda Xu, Hanwen Wang, and Guofu Cao.
\newblock {First-principle event reconstruction by time-charge readouts for
  TAO}.
\newblock {\em Eur. Phys. J. C}, 85(4):438, 2025.
\newblock [Erratum: Eur.Phys.J.C 85, 653 (2025)].

\bibitem{Li:2022gpb}
Tao Li, Yu~Chen, Shaobo Wang, Ke~Han, Heng Lin, Kaixiang Ni, and Wei Wang.
\newblock {Reconstruction of the event vertex in the PandaX-III experiment with
  convolution neural network}.
\newblock {\em JHEP}, 05:200, 2023.

\bibitem{Qian:2021vnh}
Zhen Qian et~al.
\newblock {Vertex and energy reconstruction in JUNO with machine learning
  methods}.
\newblock {\em Nucl. Instrum. Meth. A}, 1010:165527, 2021.

\bibitem{Li:2022tvg}
Zi-Yuan Li, Zhen Qian, Jie-Han He, Wei He, Cheng-Xin Wu, Xun-Ye Cai, Zheng-Yun
  You, Yu-Mei Zhang, and Wu-Ming Luo.
\newblock {Improvement of machine learning-based vertex reconstruction for
  large liquid scintillator detectors with multiple types of PMTs}.
\newblock {\em Nucl. Sci. Tech.}, 33(7):93, 2022.

\bibitem{Venettacci:2024hwy}
C.~Venettacci.
\newblock {SiPM and readout electronics for the JUNO-TAO Central Detector}.
\newblock {\em JINST}, 19(07):C07008, 2024.

\bibitem{Luo:2023inu}
Guang Luo et~al.
\newblock {Design optimization of plastic scintillators with
  wavelength-shifting fibers and silicon photomultiplier readouts in the top
  veto tracker of the JUNO-TAO experiment}.
\newblock {\em Nucl. Sci. Tech.}, 34(7):99, 2023.

\bibitem{Guan:2023uaa}
Y.~Guan, N.~Anfimov, G.~Cao, Z.~Xie, Q.~Dai, D.~Fedoseev, K.~Kuznetsova,
  A.~Rybnikov, A.~Selyunin, and A.~Sotnikov.
\newblock {Study of Silicon Photomultiplier external cross-talk}.
\newblock {\em JINST}, 19(06):P06024, 2024.

\bibitem{Zhao:2022gks}
Rong Zhao et~al.
\newblock {Afterpulse measurement of JUNO 20-inch PMTs}.
\newblock {\em Nucl. Sci. Tech.}, 34(1):12, 2023.

\bibitem{Baldi:2014kfa}
Pierre Baldi, Peter Sadowski, and Daniel Whiteson.
\newblock {Searching for Exotic Particles in High-Energy Physics with Deep
  Learning}.
\newblock {\em Nature Commun.}, 5:4308, 2014.

\bibitem{Chan:2019fuz}
Man~Leong Chan, Ik~Siong Heng, and Chris Messenger.
\newblock {Detection and classification of supernova gravitational wave
  signals: A deep learning approach}.
\newblock {\em Phys. Rev. D}, 102(4):043022, 2020.

\bibitem{Zeng:2023att}
Yu-Da Zeng, Jun Wang, Rong Zhao, Feng-Peng An, Xiang Xiao, Yuenkeung Hor, and
  Wei Wang.
\newblock {Decomposition of fissile isotope antineutrino spectra using
  convolutional neural network}.
\newblock {\em Nucl. Sci. Tech.}, 34(5):79, 2023.

\bibitem{Chen:2024vdq}
Jian Chen, Jun Wang, Wei Wang, and Yuehuan Wei.
\newblock {Extraction of fissile isotope antineutrino spectra using feedforward
  neural network}.
\newblock {\em Nucl. Sci. Tech.}, 36(10):11, 2025.

\bibitem{Psihas:2019ksa}
F.~Psihas, E.~Niner, M.~Groh, R.~Murphy, A.~Aurisano, A.~Himmel, K.~Lang, M.~D.
  Messier, A.~Radovic, and A.~Sousa.
\newblock {Context-Enriched Identification of Particles with a Convolutional
  Network for Neutrino Events}.
\newblock {\em Phys. Rev. D}, 100(7):073005, 2019.

\bibitem{Lee:2022kdn}
Kyle Lee, James Mulligan, Mateusz P\l{}osko\'n, Felix Ringer, and Feng Yuan.
\newblock {Machine learning-based jet and event classification at the
  Electron-Ion Collider with applications to hadron structure and spin
  physics}.
\newblock {\em JHEP}, 03:085, 2023.

\bibitem{Simonyan:2014cmh}
Karen Simonyan and Andrew Zisserman.
\newblock {Very Deep Convolutional Networks for Large-Scale Image Recognition}.
\newblock 9 2014.

\bibitem{Chen:2023xhj}
Guo-Ming Chen, Xin Zhang, Ze-Yuan Yu, Si-Yuan Zhang, Yu~Xu, Wen-Jie Wu,
  Yao-Guang Wang, and Yong-Bo Huang.
\newblock {Discrimination of pp solar neutrinos and $^{14}$C double pile-up
  events in a large-scale LS detector}.
\newblock {\em Nucl. Sci. Tech.}, 34(9):137, 2023.

\bibitem{DiCroce:2024dfy}
Davide Di~Croce, Massimo Giovannozzi, Carlo~Emilio Montanari, Tatiana Pieloni,
  Stefano Redaelli, and Frederik~F. Van~der Veken.
\newblock {Assessing the Performance of Deep Learning Predictions for Dynamic
  Aperture of a Hadron Circular Particle Accelerator}.
\newblock {\em Instruments}, 8(4):50, 2024.

\bibitem{Tingey:2022evd}
Josh Tingey et~al.
\newblock {Neutrino characterisation using convolutional neural networks in
  CHIPS water Cherenkov detectors}.
\newblock {\em JINST}, 18(06):P06032, 2023.

\bibitem{Girshick:2015ICCV}
Ross Girshick.
\newblock Fast r-cnn.
\newblock In {\em Proceedings of the 2015 IEEE International Conference on
  Computer Vision (ICCV)}, ICCV '15, page 1440–1448, USA, 2015. IEEE Computer
  Society.

\bibitem{Kingma:2014vow}
Diederik~P. Kingma and Jimmy Ba.
\newblock {Adam: A Method for Stochastic Optimization}.
\newblock 12 2014.

\bibitem{smith:2018scv}
Leslie~N. Smith and Nicholay Topin.
\newblock Super-convergence: Very fast training of neural networks using large
  learning rates, 2018.

\bibitem{he2015:drl}
Kaiming He, Xiangyu Zhang, Shaoqing Ren, and Jian Sun.
\newblock Deep residual learning for image recognition, 2015.

\end{thebibliography}

\end{document}